\documentclass[sigconf]{acmart}
\AtBeginDocument{%
  \providecommand\BibTeX{{%
    \normalfont B\kern-0.5em{\scshape i\kern-0.25em b}\kern-0.8em\TeX}}}

\usepackage{xspace}
\usepackage{shortcuts}
\usepackage{algorithm,algcompatible}
\usepackage{slashbox}
\usepackage{multirow}
\usepackage[export]{adjustbox}
\usepackage{seqsplit}
\usepackage{tikz}
\usepackage{amsmath}
\usepackage{bm}
 
\DeclareMathOperator*{\argmin}{argmin} 
\newcommand\boldblue[1]{\textcolor{blue}{\textbf{#1}}}
\newcommand\boldred[1]{\textcolor{red}{\textbf{#1}}}

\begin{document}

\title{\codename: A Reinforcement Learning Framework for Attacking Static Malware Classifiers}



\author{Wei Song}
\affiliation{%
  \institution{University of California, Riverside}
  \city{Riverside}
  \country{USA}
}
\email{wsong008@ucr.edu}

\author{Xuezixiang Li}
\affiliation{%
  \institution{University of California, Riverside}
  \city{Riverside}
  \country{USA}
}
\email{xli287@ucr.edu}

\author{Sadia Afroz}
\affiliation{%
  \institution{Avast, ICSI}
  \city{San Francisco}
  \country{USA}
}
\email{sadia.afroz@avast.com}
\email{sadia@icsi.berkeley.edu}

\author{Deepali Garg}
\affiliation{%
  \institution{Avast}
  \city{Santa Clara}
  \country{USA}
}
\email{deepali.garg@avast.com}

\author{Dmitry Kuznetsov}
\affiliation{%
  \institution{Avast}
  \city{Prague}
  \country{Czech}
}
\email{kuznetsov@avast.com}

\author{Heng Yin}
\affiliation{%
  \institution{University of California, Riverside}
  \city{Riverside}
  \country{USA}
}
\email{heng@cs.ucr.edu}




 \pagestyle{plain}

\renewcommand{\shortauthors}{Anonymous}

\begin{abstract}

Modern commercial antivirus systems increasingly rely on machine learning to keep up with the rampant inflation of new malware. However, it is well-known that machine learning models are vulnerable to adversarial examples (AEs). Previous works have shown that ML malware classifiers are fragile to the white-box adversarial attacks. However, ML models used in commercial antivirus products are usually not available to attackers and only return hard classification labels. Therefore, it is more practical to evaluate the robustness of ML models and real-world AVs in a pure black-box manner. We propose a black-box Reinforcement Learning (RL) based framework to generate AEs for PE malware classifiers and AV engines. It regards the adversarial attack problem as a multi-armed bandit problem, which finds an optimal balance between exploiting the successful patterns and exploring more varieties. Compared to other frameworks, our improvements lie in three points. 1) Limiting the exploration space by modeling the generation process as a stateless process to avoid combination explosions. 2) Due to the critical role of payload in AE generation, we design to reuse the successful payload in modeling. 3) Minimizing the changes on AE samples to correctly assign the rewards in RL learning. It also helps identify the root cause of evasions. As a result, our framework has much higher black-box evasion rates than other off-the-shelf frameworks. Results show it has over 74\%--97\% evasion rate for two state-of-the-art ML detectors and over 32\%--48\% evasion rate for commercial AVs in a pure black-box setting. We also demonstrate that the transferability of adversarial attacks among ML-based classifiers is higher than the attack transferability between purely ML-based and commercial AVs.

\end{abstract}



\maketitle

\section{Introduction}
\label{sec:intro} 

Malware attacks continue to be one of the most pressing security issues users face today. Recent research showed that during the first nine months of 2019, at least 7.2 billion malware attacks and 151.9 million ransomware attacks have been reported.\footnote{\url{https://www.msspalert.com/cybersecurity-research/sonicwall-research-malware-attacks-2019/}} The attack rate hit a new high with the COVID-19 pandemic.\footnote{\url{https://labs.bitdefender.com/2020/04/coronavirus-themed-threat-reports-havent-flattened-the-curve/}}
The traditional signature-based methods cannot keep up with this rampant inflation of novel malware. Hence commercial antivirus companies started using machine learning~\cite{avast_link,microsoft_link}. 
Machine-learning-based detectors are scalable and efficient at protecting against the huge influx of malware, which is why since the first paper in 2001 on detecting malware using machine learning~\cite{schultz2000data}, there has been an explosion of academic research papers on predicting malicious content using machine learning, many of them flaunting high accuracy and being able to detect new malware unseen during training ~\cite{malconv, ember, dahl2013large, rieck2011automatic, saxe2015deep}. On the other hand, research has also demonstrated that machine-learning-based detectors can be easily evaded by making even trivial changes to malware~\cite{android1,android2,android3,android4,android5,android6,android7,android8,chen2017adversarialEISI,ehteshamifar2019easy,xu2016automatically,maiorca2018towards,pe1,pe2,pe4,pe5,pe6,pe7,pe9,pe10,pe11,pe12,pe13,pe14,pe15,pe16,pe17,pe18,pe19,pierazzi2019intriguing,fass2019hidenoseek,yang2017malware,demontis2019adversarial,maiorca2015stealth,castro2019aimed}. Even commercial antivirus systems, such as Cylance, have been shown to be susceptible to trivial adversarial attacks~\cite{cylance}.

Since 2014, there have been more than 1400 papers on adversarial attacks and defense\footnote{\url{https://nicholas.carlini.com/writing/2019/all-adversarial-example-papers.html}}. However, these works mainly focus on the image domain.
The adversarial attacks on malware samples are different from attacks in the image domain. For images, adversaries can alter the value of any pixel, as long as the changes are bounded with a $L_p$-norm.
But for malware samples, a one-byte change can break the format of a valid PE, or break the original malicious functionality. This is why adversaries usually do not directly modify the raw bytes of the PE file.
Instead, they construct a set of actions. Each action can transform the malware sample without breaking the original functionality. For example, the action could be adding a new redundant section (adding a new entry in the section table and appending the section content at the end). Then, the adversarial example generation problem is transformed into finding correct actions and corresponding contents that lead to misclassification.

Adversarial attacks against static malware classifiers are not new. Researchers have proposed a variety of techniques to generate evasive samples (the terms ``evasive samples'' and ``adversarial examples'' are used exchangeably in this paper), including genetic programming~\cite{xu2016automatically,Demetrio2020FunctionalitypreservingBO}, Monte Carlo tree search~\cite{quiring2019misleading}, and deep Q-learning~\cite{pe14}. Although some of these attempts~\cite{xu2016automatically,quiring2019misleading} are dealing with PDF malware and source code authorship respectively, the general algorithms can be applied to PE malware.

Although these techniques have been demonstrated to be effective, we have identified several limitations. First, the existing techniques model the AE generation in a stateful manner, meaning that the actions depend on one another. While this modeling is general, it is usually hard to train a stateful model given that the search space is large. We observe that many actions for transforming PE malware are independent. Therefore we choose a stateless modeling approach, which can significantly reduce the learning difficulty and result in more productive AE generation.
Second, most of the existing techniques only learn a decision-making policy that decides what action to take in the next step and randomly picks content if needed. For instance, when adding a new section, it will fill the new section with random content. However, contents are as important as actions. If content associated with the action has proved to be useful in one AE, the same action-content pair will likely be useful for some other samples as well. So we should model action and its content as an integral unit. 
Third, when an AE is successfully generated, these techniques will assign rewards to all the actions involved. In our evaluation, we observe that only a small number of (mostly one or two) actions are essential and the rest are redundant. Assigning rewards to these redundant actions will confuse the learning process. 

Based on these insights, we propose an open-source reinforcement learning framework, called \codename\footnote{\url{https://github.com/weisong-ucr/MAB-malware.git}}, to generate AEs for PE malware. We model this problem as a classic multi-armed bandit (MAB) problem, by treating each action-content pair as an independent slot machine. We model each machine's reward as a Beta distribution and use Thompson sampling to select the next action and content, striking a balance between exploitation and exploration. We devise an action minimization process, which minimizes an AE by removing redundant actions and further reducing essential actions into even smaller actions (called micro-actions). We then assign rewards only to these essential micro-actions. This minimization process also helps interpret the root cause of evasions.

In summary, the contributions of this paper are as follows: 
\begin{itemize}
    \item We examine the existing algorithms in blackbox AE generation and provide key insights for stateful vs. stateless modeling, content-aware vs. content-agnostic modeling, and redundant vs. essential actions.

    \item We argue that a stateless and content-aware modeling is more suitable for generating adversarial PE malware, and an action minimization process is essential. 

    \item To meet these design choices, we propose and implement a novel MAB-based reinforcement learning framework for generating adversarial PE malware.
    
    \item We conduct an extensive evaluation using 5000 PE malware samples on two popular machine learning models and three commercial AV engines. \codename can achieve a very high evasion rate (over 75\%) for machine learning models and outperform the existing blackbox AE generation algorithms by large margins. It also shows a noticeable improvement in commercial AV engines.
    
    \item Based on our action minimization, we further look into the root cause of these evasions. Our experiment results suggest the static classifiers in the commercial AV engines are vulnerable to trivial changes to a malware sample. We also demonstrate that the transferability of adversarial attacks among ML-based classifiers is high (over 80\%) but low (less than 7\%) between purely ML-based and commercial AVs.
    
\end{itemize} 

To facilitate the follow-up research on this topic, we plan to share our framework as well as the dataset of adversarial malware samples with researchers upon request.
\section{Problem}
\label{sec:problem}
\subsection{Threat Model}

We follow the study by Carlini et al.~\cite{carlini2019evaluating} to describe our threat model, from three aspects: adversarial goal, adversarial capabilities, and adversarial knowledge.

\paragraph{Adversarial Goal.}
The adversary's goal is to manipulate malware samples to evade the detection of static PE malware classifiers. Other types of malware like PDF malware or Android malware are not within the scope of this study. This is an untargeted attack because we only consider a binary classification (benign or malicious) not specific malware families in this classification task and we are only interested in causing the malicious samples to be classified as benign.

\paragraph{Adversarial Capabilities.}
In this work, we assume that the adversary does not have access to the training phase of the malware classifiers. For instance, the adversary cannot inject poisonous data into the training dataset.

Also, the adversary cannot arbitrarily change the input data. In most scenarios of adversarial attacks, such as image recognition, the adversary is required to make only ``small'' changes to the original sample to keep the manipulation visually imperceptible. However, when attacking malware classification, the restriction is not on the number or size of changes, but on the preservation of malicious functionality. If ``small'' changes on a malware sample indeed confuse a malware classifier but prevent the malware from acting maliciously, this manipulation is not considered successful. 

\paragraph{Adversarial Knowledge.}
Based on the knowledge an adversary can obtain, an attack can be divided into two types: 1) whitebox attacks where the adversary has unlimited access to the model; and 2) blackbox attacks where the adversary has no knowledge about the model and can obtain the classification results only through a limited number of attempts. A classification result can be a score or simply a label.

In this work, we consider an adversary with only blackbox access. The adversary does not know anything about the internals of the deployed classifiers, can perform a limited number of attempts to the classifiers, and can observe the classifiers' actions when the samples are considered malicious.

\subsection{Problem Definition}

In this paper, we focus on three state-of-the-art machine learning classifiers and the static classifiers of 3 top commercial antivirus products.


We aim to automatically generate adversarial examples for malware classifiers and explain the root cause of the evasions. The problem can be split into two sub-problems: adversarial example generation and feature interpretation.


We aim to manipulate a malware sample such that malware classifiers misclassify it as a benignware, and do not break its malicious functionalities. For whitebox attacks in the image domain, changes to original images are bounded with $L_2$ and $L_\infty$ norms. It ensures that the pixel changes are imperceptible to humans. 
However, in the malware classification domain, as long as the malware behaviors remain the same, it is unlikely for normal users to notice the differences. That's why previous blackbox attacks~\cite{ceschin2019shallow,fleshman2018study,pe14} on malware do not try to minimize changes when generating AEs. However, we find that the minimal change requirement is still crucial for three main reasons: 1) it reveals which actions and the corresponding payloads are essential to generate evasive samples that can be applied to other samples to create successfully evasive samples; 2) it unveils which feature changes caused the evasion to ensure that the classifier does not rely on superficial features; 3) it reduces the chance of creating broken binaries. In the blackbox setting, instead of minimizing added noises in feature space, we minimize action sequences applied to generate AEs. It includes removing redundant actions and replacing actions that cause large changes to the features used for detection.

Let $\mathcal{X}$ be a malware dataset, $f$ be a malware classifier that maps a sample $x \in \mathcal{X}$ to a classification label $y \in \{0,1\}$ (0 represents benign, 1 represents malicious). We implement an action set $\mathcal{A}=\{a_1, a_2, \dots\, a_n\}$ that can be used to perturb malware samples.
We define an objective function for adversarial example generation in (\ref{eq:loss}). An adversarial example $x'=t(x)$ is generated by applying a transformation function $t$, which is a sequence of actions sampled from set $\mathcal{A}$. $\mathcal{L}(f(t(x)), \bar{y})$ measures the difference between the predicted label of $f(t(x))$ and benign label $\bar{y}$. The transformation function $t$ subjects to the constraint that $t(x)$ does not change the functionality of $x$, i.e. the functionality difference $\delta(x,t(x))$ before and after transformation equals to 0.

\begin{equation}
\label{eq:loss}
\begin{split}
    \argmin_{t}\mathcal{L}(f(t(x)),\bar{y}),\\ 
	s.t.\enspace\delta(x,t(x)) = 0\\
	y \neq \bar{y}
\end{split}
\end{equation}

\section{Motivation}

In this section, we first discuss the existing techniques on AE generation and their limitations, and then we present our insights that motivate our MAB-based approach.

\subsection{Existing Techniques on AE Generation}

\paragraph{Deep Q-learning.}
Anderson et al.~\cite{pe14} propose to apply deep reinforcement learning (RL) to generate AE for PE malware to bypass machine learning models. They first define a set of actions (file mutations), including changing PE headers, appending overlay bytes, packing, and unpacking. Then the agent selects the next action based on a policy and an environmental state. When an evasive sample is generated, all applied actions (including early actions that produce no immediate reward) get promoted for a given state.

\paragraph{Genetic Programming.} Demetrio et al.~\cite{Demetrio2020FunctionalitypreservingBO} propose a genetic programming-based approach to generate AEs of PE malware in a black-box attack manner. It formalizes the problem as a constrained minimization problem, to trade-off between the probability of evasion and injected payload size. It first extracts some benign content from benign binaries, and then appends a random fraction of them at the end of target malware samples. The fitness function is defined as the sum of confidences score and injected payload size. In each iteration, it selects variants with the lowest fitness score. It iterates over three steps: selection, cross-over, and mutation. Cross-over and mutation guarantee that the new population is sufficiently different from the previous variants.
Another paper from Xu et al.~\cite{xu2016automatically} also uses a genetic programming-based approach to generate adversarial PDF malware. They first define a set of actions, called mutation operators, which include deleting, inserting, and replacing an object in a PDF tree structure. It uses the confidence score as the fitness score and also conducts selection, mutation, but without cross-over. Different from Demetrio et al.'s work~\cite{Demetrio2020FunctionalitypreservingBO}, their approach saves and reuses the successful traces for new samples. The trace contains the sequence of actions and the content used by each action.

\paragraph{Monte Carlo Tree Search.}
Quiring et al.~\cite{quiring2019misleading} propose a Monte Carlo Tree Search (MCTS) based approach to mislead the classification of source code authorship. Although this work is not in the malware classification domain, the high-level idea is applicable. They define a set of actions (code transformation) for changing stylistic patterns. Then they create a Monte Carlo search tree, in which each node represents a variant of the code and each edge represents an action. 
In each iteration, the node with the highest average classifier score is selected. From that node, the framework generates a set of transformation sequences and propagates the observed classifier scores from the leaf node back to the root node. The process repeats until it generates an evasive sample. 
Then the task of AE generation is converted to a path search problem. The goal is to find a path on the tree that leads to misclassification.


\subsection{Our Insights} \label{subsec:insights}

While these existing techniques have demonstrated their effectiveness more or less, we observe several key insights, which can motivate us to develop a better technique for AE generation. 

\paragraph{Stateful vs. Stateless Modeling.}
Existing techniques model the AE generation in a stateful manner. They aim to find the best state in each iteration, select the best action in the current state according to the policy, and transform the sample to enter the next state. In other words, they try to find an optimal path of states that leads to evasion. Stateful modeling is necessary for hard tasks, such as AlphaGo and video-game-playing problems. The optimal path to reach success is often very deep. It also means that it is usually difficult and time-consuming to train a stateful model. 

Our key insight is that it is not necessary to model the problem of AE generation for PE malware in a stateful manner. Since it is generally hard to manipulate PE files without breaking their functionalities, the existing actions are rather coarse-grained and mostly independent. For instance, actions like removing debug information, section rename, section add, etc. can be applied in any order. The resulting binary file is the same regardless of the order in which these actions are applied. Of course, there are cases where actions may depend on one another. For instance, when adding multiple sections to a PE file, the order matters. Even in these cases, we argue that the dependency between these actions may not be very strong. According to our evaluation results for machine learning models and AV engines in Section~\ref{sec:explanation}, oftentimes only one or two actions are needed to generate an AE, after removing the unnecessary actions. It means that the dependency between actions (if it exists) is rather weak, at least for the machine learning models and AV engines we evaluated.

Based on this insight, we believe that stateless modeling of AE generation for PE malware is reasonable. Compared to stateful modeling, stateless modeling would treat each action independently, allowing a faster learning process and more productive AE generation. To this end,  we propose to utilize a classic reinforcement learning model, multi-armed bandit~\cite{MAB}, to model each action as an independent slot machine. It estimates the probability of each action being evasive over limited trials and leverages the estimated probabilities to select the best actions to maximize the overall reward: generating as many AEs as possible in a limited number of trials.

\paragraph{Content Modeling.}
Many actions used for manipulating PE need to be associate with some contents. For instance, when adding a new section, we need to specify what content to be filled in that section, and when renaming a section, we need to provide a new section name. Our second insight is that these contents are as important as the actions. If content associated with one action has proved to be useful in one AE, the same action-content pair will likely be useful in the future too.

Unfortunately, the existing techniques mentioned above (except the one by Xu et al~\cite{xu2016automatically}) do not take contents into account. They only learn a decision-making policy to decide what action to take in the next step and take random content if required. For example, if the next action to take is ``Section Add'' according to the policy, they will fill the new section with random content (from a pool of data). Xu et al.~\cite{xu2016automatically} indeed take contents into account, but they do not make the best use of it (only reuse traces for the first generation). Our MAB-based framework treats an action-content pair as an integral unit (a slot machine) for modeling. If a new action-content pair is discovered to be useful to generate an adversarial example, it will be saved as a new slot machine, and put into the machine pool. When other samples select that machine again, the same content will be reuse again.


\paragraph{Precise Reward Assignment.}
Reward assignment is essential to all the existing AE generation techniques described above. When an AE is successfully generated, a positive reward is assigned to the corresponding state or the corresponding sequence of actions. As discussed earlier, not all actions are essential to the generation of this AE. According to our evaluation in Section~\ref{sec:explanation}, in most cases, only one or two actions are essential and all the rest are unnecessary. Therefore, assigning rewards to all the actions involved in an AE generation will lead to a less accurate reinforcement learning model. Instead, it is better to remove the unnecessary actions, and only assign rewards to the essential actions. Hence, our third insight is that we should precisely assign rewards only to the essential actions.


\begin{figure}[thb]
    \centering
    \includegraphics[width=0.9\linewidth]{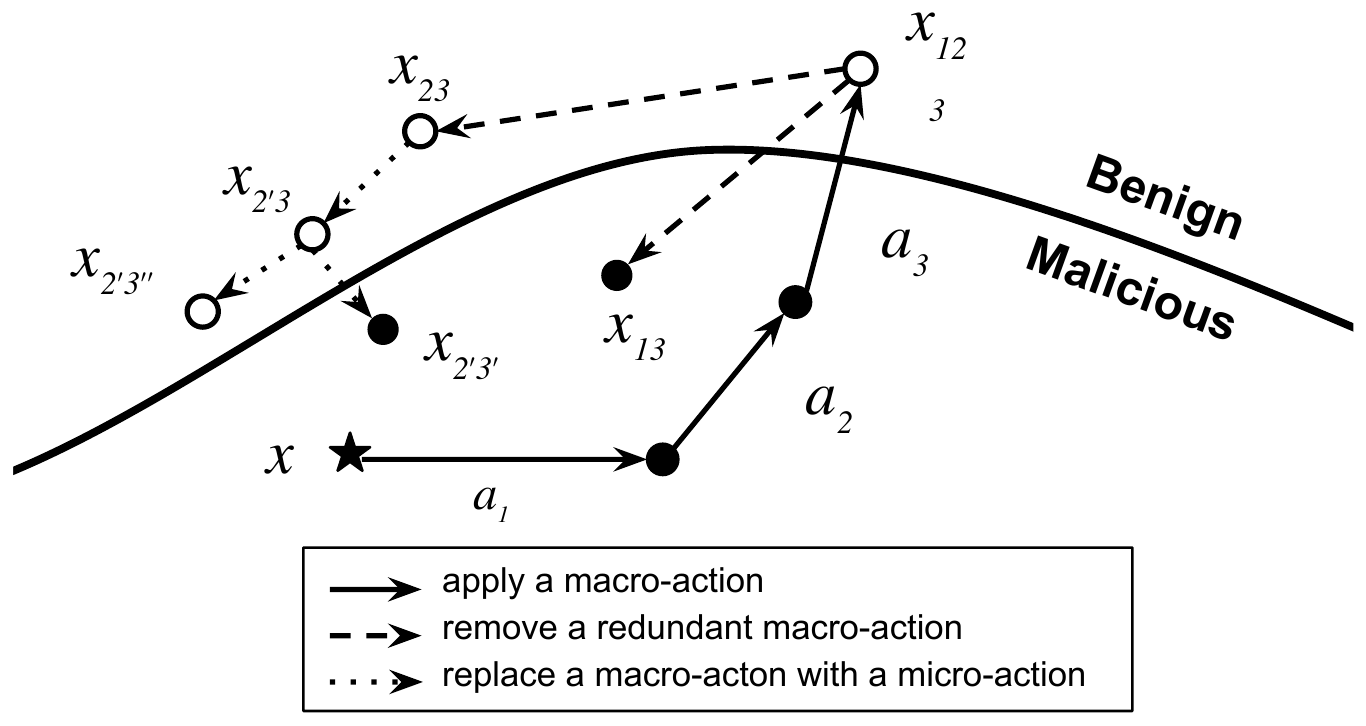}
    \caption{An example of action minimization.}
    \label{fig:minimization}
\end{figure}

To identify the essential actions for an AE, we propose a minimization process. As shown in Figure~\ref{fig:minimization}, the original malware sample $x$ resides in the malicious region of the feature space. We perform a sequence of single actions $a_1$, $a_2$ and $a_3$ until the generated sample $x_{123}$ successfully reaches the benign region. $x_{123}$ is an adversarial example. In the minimization phase, first, we remove useless actions. The action $a_2$ is essential, because by removing action $a_2$, the generated sample $x_{13}$ is no longer evasive. The action $a_1$ is useless because by removing action $a_1$, the generated sample $x_{23}$ has no effect in the classifier's decision. Then we disentangle these actions into micro ones (i.e., actions that cause smaller changes). $a_2$ can be replaced with micro-actions $a_{2}'$. Action $a_3$ can be replaced with micro-actions $a_{3}'$ or $a_{3}''$. We generate three samples $x_{2'3}$, $x_{2'3'}$ and $x_{2'3''}$. Finally, we have an adversarial sample $x_{2'3''}$ with a minimized action sequence $(a_2',a_3'')$. So a positive reward can be precisely assigned to these essential actions $a_2'$ and $a_3''$.

\section{MAB-Malware}
\label{sec:methodology}

\begin{figure*}[thb]
    \centering
    \includegraphics[width=\linewidth]{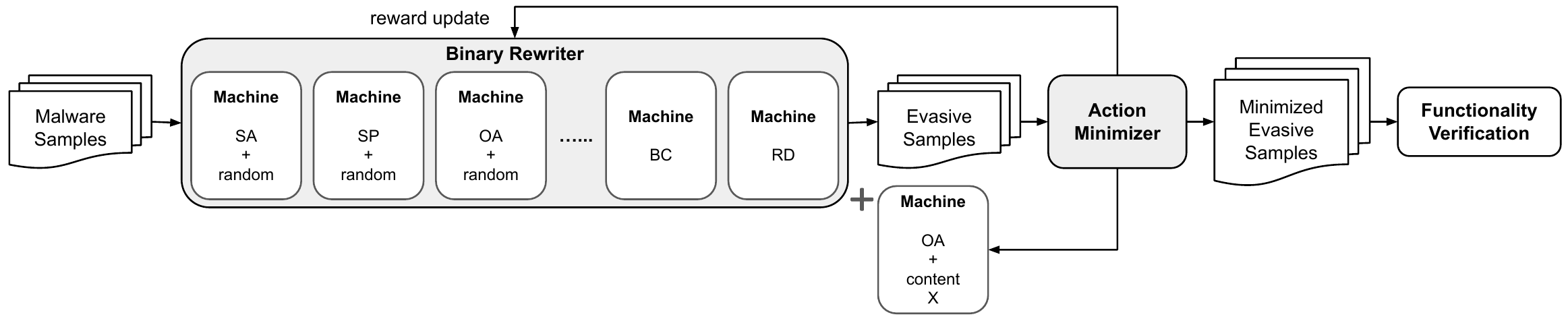}
    \caption{Workflow}
    \label{fig:overview}
\end{figure*}

\subsection{The Framework}
The workflow of our framework MAB-Malware is shown in Figure~\ref{fig:overview}. It consists of two main modules: the Binary Rewriter and the Action Minimizer. The Binary Rewriter utilizes Thompson sampling to select action sequence $t$ from the action set $\mathcal{A}$ and rewrites original malware sample $x$ to generate variants $x'=t(x)$. If $x'$ can evade the detection of the target classifier, the Action Minimizer continues to remove redundant actions from $t$ to generate minimized evasive samples $x'_{min}$ and infer the root cause of the evasion. At last, the framework verifies the functionality of $x'_{min}$. If the behavior of it is changed, the original sample $x$ is put back into the working queue. Otherwise, $x'_{min}$ is returned.

\subsection{Binary Rewriter}
\subsubsection{Action Set and Features}
\label{sec:action_set_and_features}

\begin{table*}[htbp]
\caption{Action Set.}
\begin{center}
{\small
\begin{tabular}{l|l|l|l}
\hline
\textbf{Type}   & 
\textbf{Abbr} & \textbf{Name} & \textbf{Description}                                          \\ \hline
\multirow{8}{*}{Macro} & 
    OA   & Overlay Append              & Appends benign contents at the end of a binary                 \\
&    SP   & Section Append              & Appends random bytes to the unused space at the end of a section.                 \\
&    SA   & Section Add                 & Adds a new section with benign contents. \\
&    SR   & Section Rename              & Change the section name to a name in benign binaries.                           \\
&    RC   & Remove Certificate          & Zero out the signed certificate of a binary.                                      \\
&    RD   & Remove Debug                & Zero out the debug information in a binary.                                       \\
&    BC   & Break Checksum              & Zero out the checksum value in the optional header.                               \\
&    CR   & Code Randomization          & Replace instruction sequence with semantically equivalent one                     \\
\hline
\multirow{4}{*}{Micro} & 
    OA1  & Overlay Append 1 Byte      & Appends 1 byte at the end of a binary                                              \\
&    SP1  & Section Append 1 Byte      & Appends 1 byte to the unused space at the end of a section.                        \\
&    SA1  & Section Add 1 Byte         & Adds a new section with 1 byte content.                                            \\
&    SR1  & Section Rename 1 Byte      & Change 1 byte of a section name.                                                   \\
&    CP1  & Code Section Append 1 Byte & Appends 1 byte to the unused space at the end of the code section.                 \\
\hline       
\end{tabular}
\label{tab:action_set}
}
\end{center}
\end{table*}

\begin{table*}[th]
\caption{Affected Features by Actions.}
\begin{center}
\small
\begin{tabular}{|l|l|c|c|c|c|c|c|c|c|c|c|c|c|c|}
\hline
\multicolumn{2}{|c|}{}                                      &CR &OA &SP &SA &SR &RC &RD &BC &OA1 &SP1 &SA1 &SR1 &CP1\\ \hline
\multirow{2}{*}{\vtop{\hbox{\strut Hash-Based}\hbox{\strut Signatures}}}   & $F_1$: File Hash 
                                                            & \checkmark  & \checkmark & \checkmark & \checkmark & \checkmark & \checkmark & \checkmark & \checkmark & \checkmark  & \checkmark  & \checkmark  & \checkmark  &\checkmark\\ \cline{2-15}
                        & $F_2$: Section Hash               & \checkmark &   & \checkmark &   &   &   & \checkmark &   &    & \checkmark  &    &    &\checkmark\\ \hline
\multirow{6}{*}{\vtop{\hbox{\strut Rule-based}\hbox{\strut Signatures}}}   & $F_3$: Section Count               
                                                            &   &   &   & \checkmark &   &   &   &   &    &    & \checkmark  &    &\\ \cline{2-15}
                        & $F_4$: Section Name               &   &   &   &   & \checkmark &   &   &   &    &    &    & \checkmark  &\\ \cline{2-15}
                        & $F_5$: Section Padding            &   &   & \checkmark &   &   &   &   &   &    &    &    &    &\\ \cline{2-15}
                        & $F_6$: Debug Info                 &   &   &   &   &   &   & \checkmark &   &    &    &    &    &\\ \cline{2-15}
                        & $F_7$: Checksum                   &   &   &   &   &   &   &   & \checkmark &    &    &    &    &\\ \cline{2-15}
                        & $F_8$: Certificate                &   &   &   &   &   & \checkmark &   &   &    &    &    &    &\\ \cline{2-15}
                        & $F_9$: Code Sequence              & \checkmark &   &   &   &   &   &   &   &    &    &    &    &\\ \hline
Data Distribution       & $F_{10}$: Data Distribution       &   & \checkmark &   & \checkmark &   &   &   &   &    &    &    &    &\\ \hline
\end{tabular}
\label{tab:action_feature_mapping}
\end{center}
\end{table*}

In Table~\ref{tab:action_set}, we define the actions that can be applied to malware to create adversarial samples. Each action manipulates a set of features that a classifier may use to detect malware. Examination of open-source malware detectors such as EMBER is used to hypothesize about the features that might be used in commercial AVs.
These features are categorized into three categories: hash-based signatures (file hash and section hash), rule-based signatures (section count, section name, section padding, debug info, checksum, certificate, and code sequences), and data distribution based features (byte histogram, and byte entropy histogram).

\smallskip\noindent\textbf{Macro-actions.}
We implement most actions proposed by Anderson et al.~\cite{pe14} using the pefile library and fix many corner cases that may break the functionality.
We also adopt a code randomization action (CR) from Pappas et al.~\cite{ORP}. It is a defense method originally proposed to prevent Return Oriented Programming (ROP) attacks.

\smallskip\noindent\textbf{Micro-actions.} 
If an action $a$ changes feature set $F=\{f_1, f_2, \dots\, f_k\}$ of a malware sample, then another action that changes only a subset of $F$ is a micro-action of $a$. We implement 5 micro-actions for macro-action OA, SP, SA, SR, and CR. Table~\ref{tab:action_feature_mapping} shows all the actions used in our framework and the corresponding affected features for each action. Consider SP action (append content at the end of a section) as an example: by looking up Table~\ref{tab:action_feature_mapping}, we find that SP action affects features File Hash ($F_1$), Section Hash ($F_2$) and Rule-Based Signatures in the section padding data($F_5$). $F_1$, $F_2$ are affected if any modification is made to the file and section content. $F_5$ is affected only if SP action modifies the padding content that may contain body-based signatures.  Thus, one micro action of SP is SP1 (append 1 byte at the end of a section), as shown in Table~\ref{tab:action_feature_mapping}, which does not affect $F_5$. OA1 action (Append 1 byte at the end of a binary) is also a micro action of SP since it only affects $F_1$.

\subsubsection{Multi-armed Bandit (MAB) Problem.}

In the AE generation problem, each action with certain concrete content may or may not be useful to generate an evasive sample. We consider each machine $m$ provides a random reward from a probability distribution specific to that tuple. Our objective is to maximize the sum of rewards earned through a sequence of action-content pairs. Since the trials are limited, at each trial, we need to tradeoff between "exploitation" of the tuple that has the highest expected reward and "exploration" to obtain more knowledge about the rewards of other tuples. That's why we model this problem as a multi-armed bandit (MAB) problem~\cite{MAB}. The MAB problem is a classic reinforcement learning problem. It maximizes gains by allocating limited resources to multiple competing choices, and the property of each choice is gradually learned in the process of resource allocation. Each action-content pair is considered as a machine $m$.

\subsubsection{Thompson Sampling.}

In our task, we face the delayed feedback Problem. When evaluating the static modules of commercial antivirus systems, we need to copy the generated sample to the virtual machine with antivirus and wait for the scanning result. This process takes seconds, even minutes for certain AVs. If we adopt a deterministic algorithm, such as UCB1 or Bayesian UCB, it will always select the one with the highest values before the result returns. It causes inefficient trials because of outdated information. To address this issue, we use Thompson sampling algorithm~\cite{TS}, which is more robust than deterministic algorithms in the delayed feedback environment~\cite{chapelle2011empirical}.

We assume the reward returned by each machine $m$ follows a beta distribution~\cite{beta} specific to that tuple. The beta distribution is a continuous probability distribution parameterized by two positive parameters, denoted by $\alpha$ and $\beta$, i.e. $m \sim \operatorname{Beta}(\alpha, \beta)$.

We consider each machine $m$ returns a random reward from a beta distribution specific to that tuple.
Formally, we assume the rewards for machine $m$ follows the distribution $r(a) \sim Bernoulli(\theta)$. The expected reward $\theta$ is a fixed value and unknown to the player. When pulling machine $m$, it gets 1 with probability $\theta$, and 0 with probability $1-\theta$. In the beginning, for each machine, $\alpha$=1, $\beta$=1. When an evasive sample is generated, for every action used, if it cannot be minimized by the minimizer (we will elaborate on the minimization process in the next section), we increase its $\alpha$ by 1. Otherwise, we increase its $\beta$ by 1. In other words, $\alpha$ and $\beta$ correspond to the counts of success or fail respectively.

At each action selection iteration, for each machine, we sample a value from its $Beta(\theta;\alpha, \beta)$ distribution and select the machine with the highest value as the next machine. When the $\alpha$ and $\beta$ values of a machine are small, the uncertainty of $m$ is high. Even if this average reward is lower than other machines, it still has a relatively high possibility to get a large value. In this way, new machines are selected for exploration. After several trials, the $\alpha$ and $\beta$ value of becoming large, and the uncertainty decrease. In this way, machines with high average rewards are selected for exploitation. Thompson sampling can automatically find an optimal balance between exploration and exploitation.
Because every machine has a possibility to be selected, Thompson sampling is more robust than deterministic algorithms, such as UCB1.

When an evasive sample is generated and minimized, besides update the regards for existing machines, we also need to add new machines to our machine pool. If an essential machine includes content, we can create a new machine, with $\alpha$=1, $\beta$=1. In the beginning, the new machine's uncertainty is high, and it is relatively easy to be selected. When selected, the new machine reuses the successful content.

\subsubsection{Workflow of Binary Rewriter}
Algorithm~\ref{alg:rewriter} summarizes the workflow of the Binary Rewriter. For a seed malware sample $s$, we aim to generate a minimized evasive sample as $x'_{min}$, such that it evades the target classifier ($f(x'_{min})=0$) and only change minimal features. First, we add the original machines to the machine list $\mathcal M$, the $\alpha$ value and $\beta$ value of each machine are set to 1. When these machines need content, they will select random content. In each iteration, for each machine, we sample a value based on its $\beta$ distribution, and select the machine with the highest value, apply the corresponding machine to the current sample. We apply various actions iteratively until we get an evasive sample or exceed the total number of attempts  $max\_attempt$. 
If the sample becomes evasive, we further use Acton Minimizer to remove redundant machines. If the remaining machines are new (with new content), we add a new machine $m'$. If it's an existing machine, we increase the $\alpha$ value of it and its parent machine, which has the same action but with random content. If the current machine cannot create an evasive sample, we increase its $\beta$ value.

\begin{algorithm}[H]
    \caption{Adversarial Example Generation}\label{alg:rewriter}\
    \begin{flushleft}
    \hspace*{\algorithmicindent} \textbf{Input:} malware sample $x$ \\
    \hspace*{\algorithmicindent} \textbf{Output:} minimized evasive examples $x'_{min}$
    \end{flushleft}
    \begin{algorithmic}[1]
      \STATE $\mathcal M \leftarrow initial\_actions\_set$
      \STATE $x' \leftarrow x$
      \WHILE{$max\_attempt > 0$}
        \STATE $m \leftarrow $ max(betaSampling($\mathcal M$))
        \STATE $x' \leftarrow m.$transfer$(x)$
        \IF{isEvasive($x'$) == True}
          \STATE $x'_{eva}$ = $x'$
          \STATE $x'_{min}$ = ActionMinimization($x'_{eva}$)
          \STATE $\mathcal M_{min} \leftarrow$ getMachines($x'_{min}$)
          \FORALL{$m' \in \mathcal M_{min} $}
            \IF{$m' \notin \mathcal M$}
              \STATE $\mathcal M.add((m')$
            \ELSE
              \STATE $\alpha_{m'} = \alpha_{m'} + 1$
            \ENDIF
          \ENDFOR
        \ELSE
          \STATE $\beta_{m} = \beta_{m} + 1$
        \ENDIF
        \STATE $max\_attempt = max\_attempt - 1$
      \ENDWHILE
      \STATE return $x'_{min}$
    \end{algorithmic}
\end{algorithm}

\subsection{Action Minimizer}
\label{sec:minimizer}

The Action Minimizer removes unnecessary actions and uses micro-actions to replace macro-actions, to produce a ``minimized'' evasive sample that only changes minimal features to evade.

\begin{algorithm}[H]
\caption{Action Minimization}\label{alg:minimizer}
\begin{flushleft}
\hspace*{\algorithmicindent} \textbf{Input:} evasive sample $x'_{eva}$ \\
\hspace*{\algorithmicindent} \textbf{Output:} minimized evasive sample $x'_{min}$
\end{flushleft}
\begin{algorithmic}[1]
  \STATE $x \leftarrow$ getOriginal($x'_{eva}$)
  \STATE $action\_seq \leftarrow$ getActionSeq($x'_{eva}$)
  \FORALL{$action \in action\_seq$}
    \STATE $x' = x$.apply($action\_seq - action$)
    \IF{isEvasive($x'$) == True}
      \STATE $action\_seq = action\_seq - action$
      \STATE $x'_{min} = x'$
    \ELSE
      \FORALL{$micro \in $get\_micro\_actions$(action)$}
        \STATE $x' = x$.apply($action\_seq - action + micro$)
        \IF{isEvasive($x'$) == True}
          \STATE $x'_{min} = x'$
          \STATE break
        \ENDIF
      \ENDFOR
    \ENDIF
  \ENDFOR
\STATE return $x'_{min}$
\end{algorithmic}
\end{algorithm}
  
As shown in Algorithm~\ref{alg:minimizer}, for the action sequence $q = (a_1, a_2, ...)$ of an   $q$ trying to remove the action. We apply the new sequence on the original sample $s$ to generate a new sample $x'$. If $x'$ is evasive, we consider the action $a_i$ redundant, and remove $a_i$ from the action sequence. If not, we further use micro-actions to replace the original action with minimal feature changes. For example, the SA action (add a new section) changes many features of the original binary. It creates a new entry in the section table, adds a chunk of content at the end of the file, it also changes the file hash. If we replace this action with the OA1 action (append 1-byte overlay data), and it still can evade detection, then we can use OA1 to replace SA. In this way, we generate the minimized evasive sample $x'_{min}$.

\begin{figure}[t]
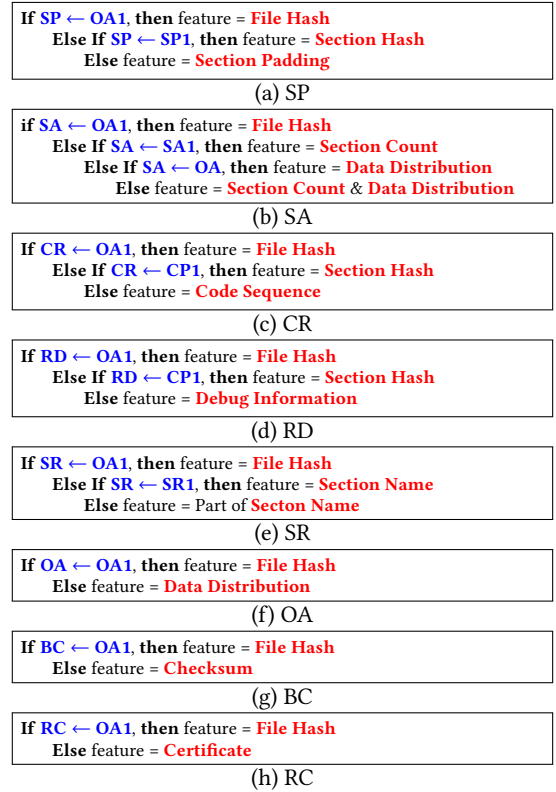

\begin{tabular}{c}
\fbox{\begin{minipage}{22em}
\footnotesize
\textbf{If} \boldblue{SP $\leftarrow$ OA1}, \textbf{then} feature = \boldred{File Hash}

\setlength\parindent{12pt}
\textbf{Else If} \boldblue{SP $\leftarrow$ SP1}, \textbf{then} feature = \boldred{Section Hash}

\setlength\parindent{24pt}
\textbf{Else} feature = \boldred{Section Padding}
\end{minipage}}
\\
(a) SP \\
\fbox{\begin{minipage}{22em}
\footnotesize
\textbf{if} \boldblue{SA $\leftarrow$ OA1}, \textbf{then} feature = \boldred{File Hash}

\setlength\parindent{12pt}
\textbf{Else If} \boldblue{SA $\leftarrow$ SA1}, \textbf{then} feature = \boldred{Section Count}

\setlength\parindent{24pt}
\textbf{Else If} \boldblue{SA $\leftarrow$ OA}, \textbf{then} feature = \boldred{Data Distribution}

\setlength\parindent{36pt}
\textbf{Else} feature = \boldred{Section Count} \& \boldred{Data Distribution}
\end{minipage}}
\\
(b) SA \\
\fbox{\begin{minipage}{22em}
\footnotesize
\textbf{If} \boldblue{CR $\leftarrow$ OA1}, \textbf{then} feature = \boldred{File Hash}

\setlength\parindent{12pt}
\textbf{Else If} \boldblue{CR $\leftarrow$ CP1}, \textbf{then} feature = \boldred{Section Hash}

\setlength\parindent{24pt}
\textbf{Else} feature = \boldred{Code Sequence}
\end{minipage}}
\\
(c) CR \\
\fbox{\begin{minipage}{22em}
\footnotesize
\textbf{If} \boldblue{RD $\leftarrow$ OA1}, \textbf{then} feature = \boldred{File Hash}

\setlength\parindent{12pt}
\textbf{Else If} \boldblue{RD $\leftarrow$ CP1}, \textbf{then} feature = \boldred{Section Hash}

\setlength\parindent{24pt}
\textbf{Else} feature = \boldred{Debug Information}
\end{minipage}}
\\
(d) RD \\
\fbox{\begin{minipage}{22em}
\footnotesize
\textbf{If} \boldblue{SR $\leftarrow$ OA1}, \textbf{then} feature = \boldred{File Hash}

\setlength\parindent{12pt}
\textbf{Else If} \boldblue{SR $\leftarrow$ SR1}, \textbf{then} feature = \boldred{Section Name}

\setlength\parindent{24pt}
\textbf{Else} feature = Part of \boldred{Secton Name}
\end{minipage}}
\\
(e) SR \\
\fbox{\begin{minipage}{22em}
\footnotesize
\textbf{If} \boldblue{OA $\leftarrow$ OA1}, \textbf{then} feature = \boldred{File Hash}

\setlength\parindent{12pt}
\textbf{Else} feature = \boldred{Data Distribution}
\end{minipage}}
\\
(f) OA \\
\fbox{\begin{minipage}{22em}
\footnotesize
\textbf{If} \boldblue{BC $\leftarrow$ OA1}, \textbf{then} feature = \boldred{File Hash}

\setlength\parindent{12pt}
\textbf{Else} feature = \boldred{Checksum}
\end{minipage}}
\\
(g) BC \\
\fbox{\begin{minipage}{22em}
\footnotesize
\textbf{If} \boldblue{RC $\leftarrow$ OA1}, \textbf{then} feature = \boldred{File Hash}

\setlength\parindent{12pt}
\textbf{Else} feature = \boldred{Certificate}
\end{minipage}}
\\
(h) RC
\end{tabular}
\caption{Decision rules are used to map actions to feature space}
\label{decision_rules}
\end{figure}

From a defender's point of view, we also would like to understand how evasion happens, where the weakest point of the classifiers is. The action minimization of evasive samples provides a good opportunity to infer that information. Figure~\ref{decision_rules} shows how we break macro-actions into micro-actions. Take the action Section Append (SP) as an example. First, by looking up Table~\ref{tab:action_feature_mapping}, SP changes feature $F= \{F_1, F_2, F_5\}$ (File Hash, Section Hash and Section padding). The actions that only change a subset of $F$ are OA1 that changes $\{F_1\}$ and SP1 that changes $\{F_1, F_2\}$. Starting from the minimum change, we try to replace SP with OA1 and check if the file is still evasive. If so, we can say the evasion is caused by breaking the file hash ($F_1$). If not, we continue to replace SP with SP1. If successful, the evasion is caused by breaking section hash ($F_2$). Otherwise, the evasion is caused by breaking signatures in section padding content ($F_5$).


\section{Evaluation}
\label{sec:evaluation}
\subsection{Experiment Setup}

\paragraph{Dataset:}
In this paper, we generate adversarial examples for Windows PE binaries. 
To ensure the executability and functionality of the generated samples, the format and constraints of PE files must remain intact.

To guarantee the quality of malware samples, we randomly select 5000 samples from VirusTotal that meet the following requirements: 1) more than 80\% antivirus engines of VirusTotal label them malicious; and 2) the execution of those samples in a Cuckoo sandbox shows malicious behavior.
Visual Basic (VB) programs are also excluded from the dataset because the IDA Pro in the implementation of code randomization~\cite{ORP} cannot generate CFG for them. These implementation issues are left as future work.

\smallskip\noindent\textbf{Setup:} 
The experiments are performed on 20 virtual machines of the Microsoft Azure cloud platform. The configuration of each virtual machine is Standard D2s v3 (2 vcpus, 8 GiB memory). All the scripts such as Binary Rewriter, Action Minimizer, and result analysis are implemented in Python 3.6.9. The Binary Rewriter also requires the pefile library and IDA Pro 6.8. For all the antivirus software under testing, free versions and default settings are used. We choose three top commercial antivirus products for blackbox testing, which are anonymized as AV1, AV2, and AV3. Each antivirus is installed on an Azure virtual machine with Windows 7 Version 6.1 Build 7601 (Service Pack 1).

To ensure the malware will not infect other machines in the network and the stability and reproducibility of our experiments, all network traffic is routed to an InetSim instance on the host machine to provide simulated network services. 


We choose the following models as our target models: 
\begin{itemize}
    \item \textbf{EMBER}~\cite{ember} is an open-source machine-learning-based classifier that uses a tree-based classifier model LightGBM to detect malware. It generates a 2350-dimensional feature vector for each sample consisting of two main types of features: raw features (e.g. ByteHistogram, ByteEntropyHistogram, Strings) and parsed features (e.g. GeneralFileInfo, HeaderFileInfo, SectionInfo, ImportsInfo, ExportsInfo). We use the implementations provided by Machine Learning Security Evasion Competition (MLSEC) 2019\cite{MLSEC2019}.
    
    \item \textbf{MalConv}~\cite{malconv} is a malware detection model that uses a neural network to learn knowledge directly on the raw bytes of malware samples. We use the model implemented in MLSEC 2019~\cite{MLSEC2019}.
    
    \item \textbf{Commercial AVs.} We also test the static classifiers of 3 top commercial antivirus systems, according to PC Magazine~\cite{pcmag}.
\end{itemize}

\subsection{Adversarial Example Generation}

\begin{figure}[t]
\begin{tabular}{c}
\includegraphics[valign=T,width=45ex]{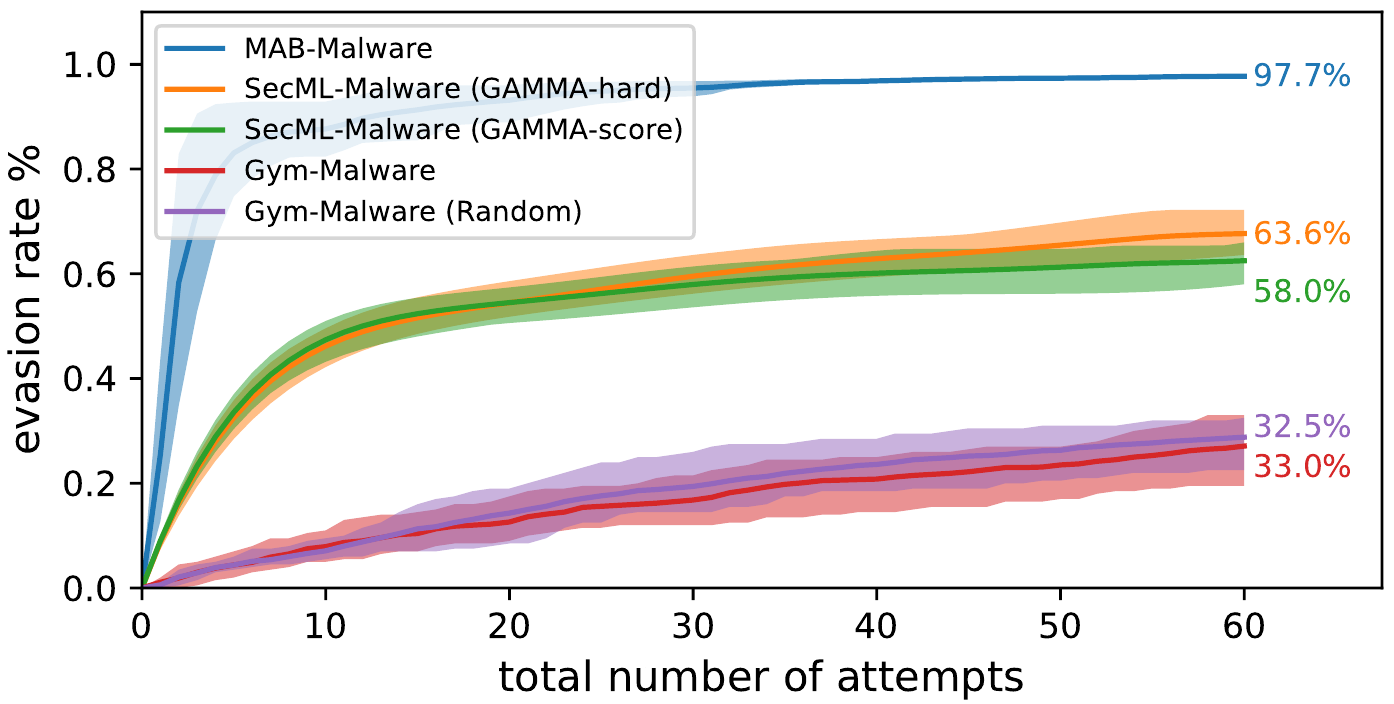} \\
(a) MalConv \\[6pt]
\includegraphics[valign=T,width=45ex]{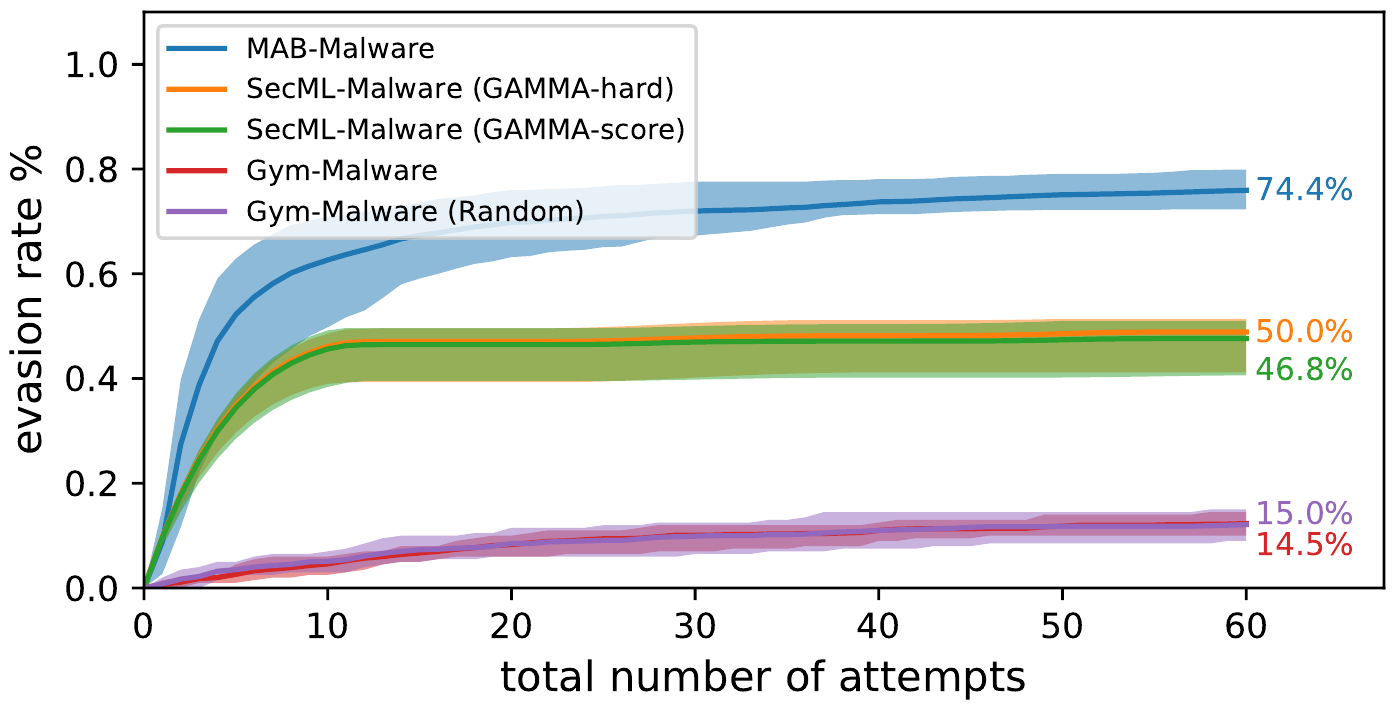}  \\
(b) EMBER \\[6pt]
\end{tabular}
\caption{Evasion Results on Different Frameworks}
\label{fig:evasion_comparison_off_shelf}
\end{figure}

\begin{figure}[t]
\begin{tabular}{c}
\includegraphics[valign=T,width=45ex]{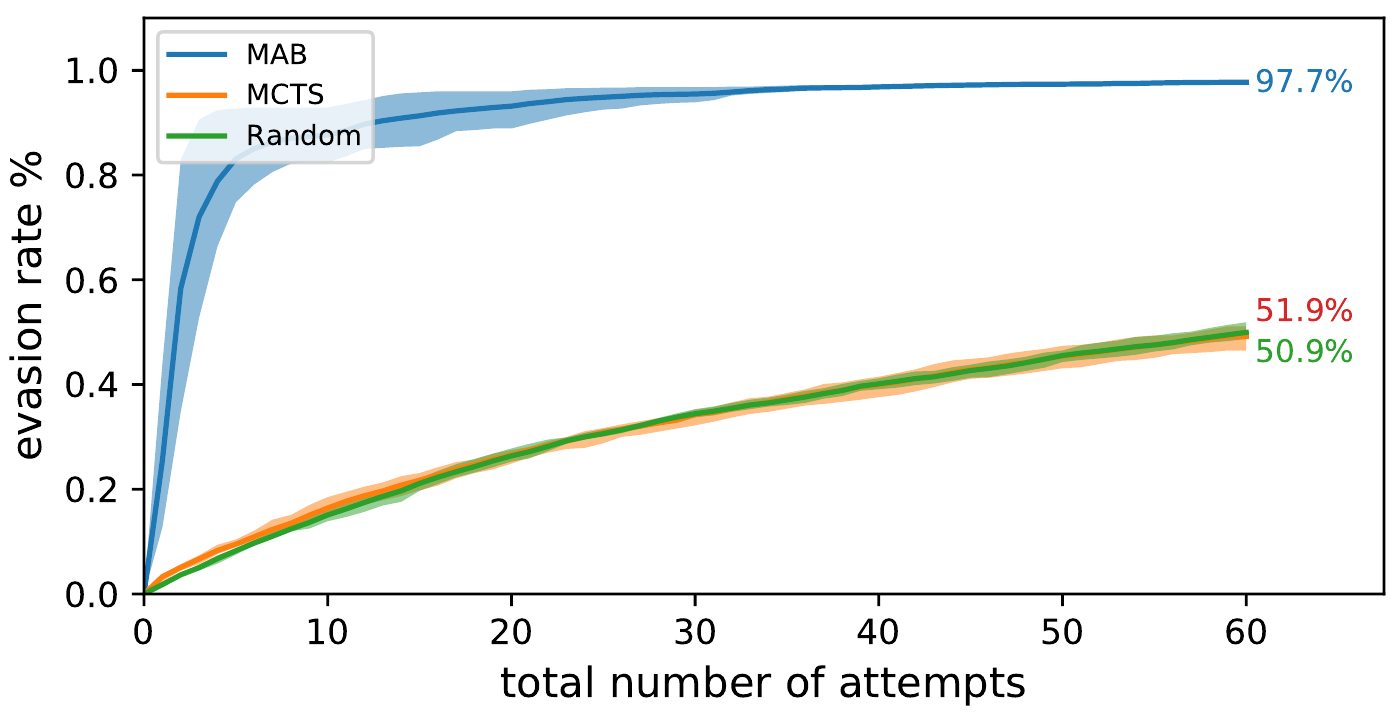} \\
(a) MalConv \\[6pt]
\includegraphics[valign=T,width=45ex]{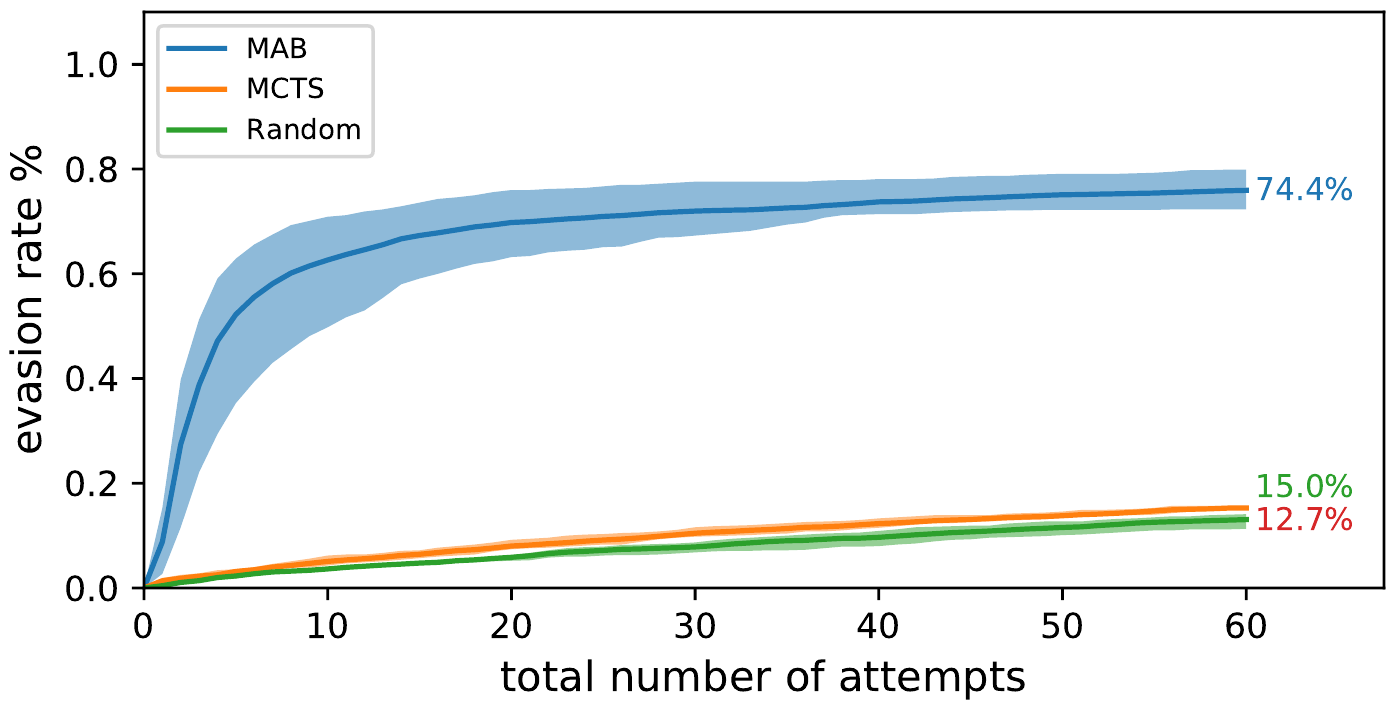}  \\
(b) EMBER \\[6pt]
\end{tabular}
\caption{Evasion Results on Different RL Algorithms}
\label{fig:evasion_comparison_MAB_actions}
\end{figure}

\begin{figure*}[t]
\begin{center}
\begin{tabular}{ccc}
\includegraphics[valign=T,width=40ex]{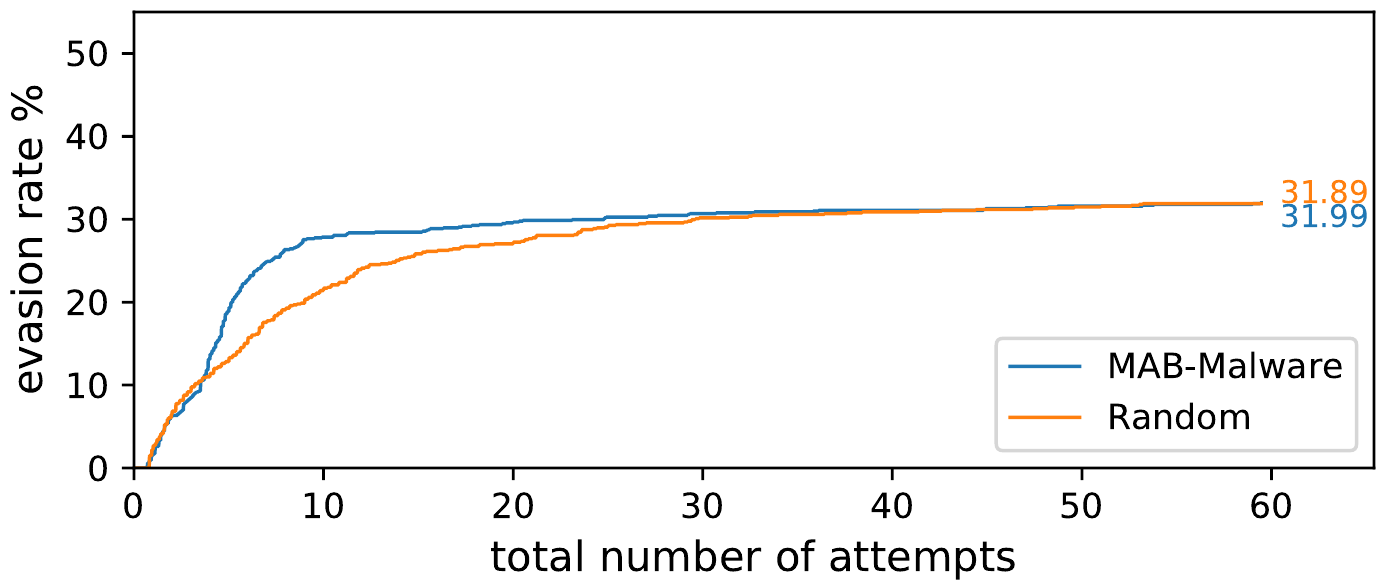} &
\includegraphics[valign=T,width=40ex]{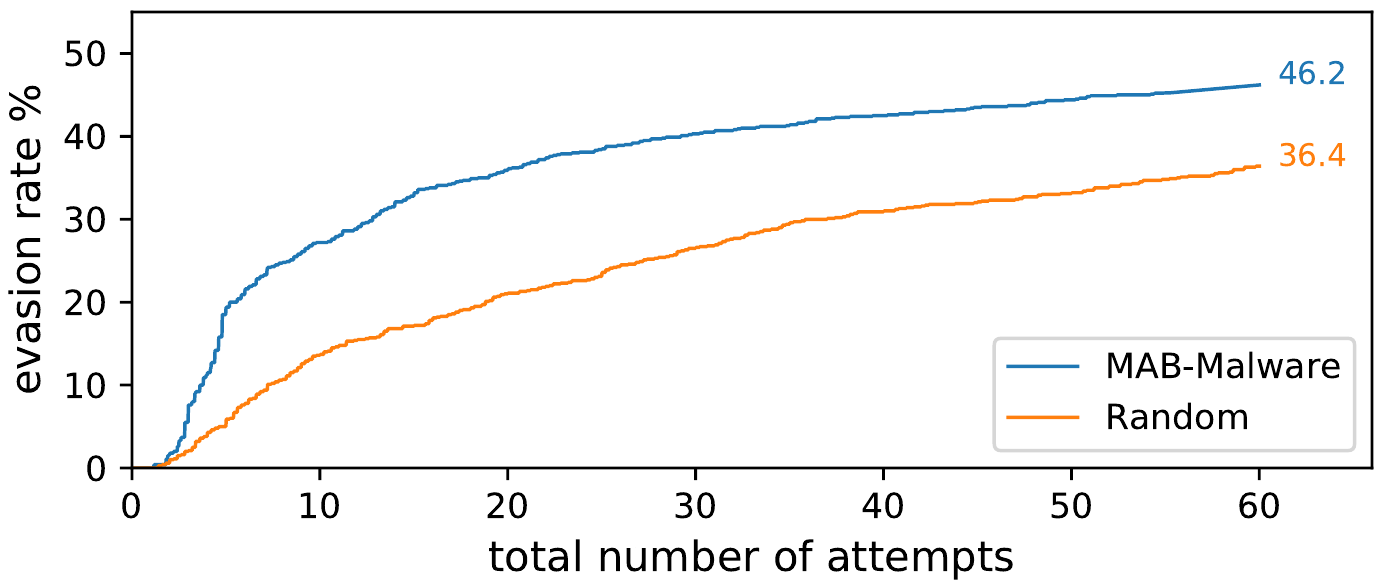} &
\includegraphics[valign=T,width=40ex]{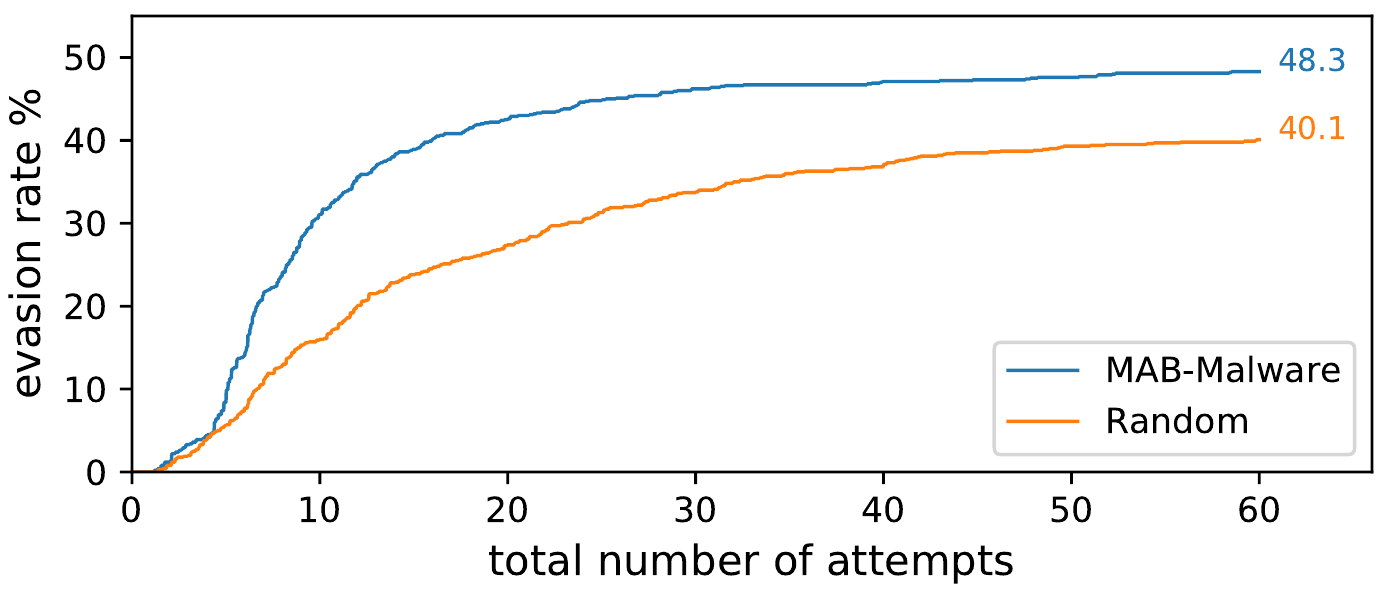} \\
(a) AV1  & (b) AV2 & (c) AV3 \\[6pt]
\end{tabular}
\caption{Evasion Results on Antivirus Engines}
\label{fig:av_result}
\end{center}
\end{figure*}

\paragraph{Comparison with Other Off-the-Shelf Frameworks.}
We compare our MAB-Malware with other two off-the-shelf attack frameworks: SecML-Malware\footnote{\url{https://github.com/zangobot/secml_malware.git}} and Gym-Malware\footnote{\url{https://github.com/endgameinc/gym-malware.git}}.
SecML-Malware is a plugin for the SecML Python library. It contains many kinds of attacks, including black-box attacks and white-box attacks. We utilize its genetic programming-based black-box attack (GAMMA) in this experiment. In the selection step, it supports both confidence score-based selection and hard label-based selection (classifiers only return benign or malicious). We assume attackers cannot get the confidence score in our threat model. So we only evaluate the hard label-based attack.
Gym-Malware is a reinforcement learning-based malware manipulation environment using OpenAI's gym. Its agents learn how to manipulate PE files to bypass AV based on a reward provided by taking specific manipulation actions. To evaluate the effect of reinforcement learning, it supports both reinforcement learning-based agent action section and random action selection.

We measure the evasion rate for two machine learning-based models, MalConv and EMBER. Evasion Rate is defined as:

\begin{equation}
\label{eq:loss}
 R_e = N_e/N_d
\end{equation}

\noindent where $N_e$ is the total number of successful evasive samples, and $N_d$ is the total number of original samples that can be detected by the target model. For a fair comparison, we use the same dataset (5000 samples from VirusTotal) and the same trained MalConv and EMBER models (from the Machine Learning Static Evasion Competition 2019~\cite{MLSEC2019}.) in all three frameworks. Each experiment runs five times to calculate the average. 

From Figure~\ref{fig:evasion_comparison_off_shelf}, we can see that our MAB-Malware framework works much better than other approaches. It can generate AEs for 97.72\% samples to evade MalConv, 74.4\% samples to evade EMBER.
The evasion rate of SecML-Malware (GAMMA-hard label) only 63.6\% and 50.0\% respectively. We can also see that even with the knowledge of the confidence score, the evasion rates are similar and sometimes even worse.
Gym-Malware has the lowest evasion rate. The evasion rate is almost identical with or without reinforcement learning. This indicates that this deep Q-learning model does not learn meaningful knowledge to guide the evasion. The reason is that the problem modeling creates an exponentially large search space. And without action minimization, the reward assignment is chaotic. Within 60 trials, it cannot explore enough in such a large space and learn meaningful policy to select the correct action and corresponding content.

\paragraph{Comparison with Other Algorithms.}
In the previous experiment, we observe that MAB-Malware generates much more adversarial examples than the other off-the-shelf frameworks. However, the action set of these three frameworks is different. SecML-Malware only utilizes benign content injection and appending. Gym-Malware's action set is much larger, including header manipulation and packing. Our action set is re-implemented using {\tt pefile} and contains more micro-actions. As a result, it is still unclear which part of our framework causes the differences: the action set, or our unique problem modeling and reinforcement learning algorithm.

So in this experiment, we only use our own action set and change the action selection algorithms. The baseline is random selection. Then we need to compare our method with the other reinforcement learning algorithms. In the experiment above, we have already shown that the Q-learning models cannot directly improve the evasion rate over random selection. In this experiment, we further implement another MCTS-based reinforcement learning algorithm. Quiring et al.~\cite{quiring2019misleading} proposed an MCTS-based approach to mislead the classification of source code authorship. Because their code cannot be directly applied to malware classifiers, we borrowed their ideas and reimplement the selection, simulation \& expansion, and backpropagation operation for our problem.

From Figure~\ref{fig:evasion_comparison_MAB_actions}, we find that in the same action set, our MAB algorithm greatly improves the evasion rate over random action selection, while the MCTS algorithm barely provides any improvement. 
Recall that unlike the existing frameworks, we model the problem differently. Existing frameworks model the AE generation in a stateful manner, and try to find an optimal path of states that leads to an evasion. When they generate an AE, they do not minimize the feature changes and assign rewards to potentially redundant actions. It makes the training difficult in a large searching space. Moreover, existing frameworks do not have a mechanism to efficiently reuse successful payloads. Whenever an action is selected, they need to start searching for the correct payload from scratch repeatedly.



\paragraph{Attacking Commercial Antivirus.} We also test our framework on three commercial antivirus engines. Since the throughputs of commercial AV engines are much lower than machine learning classifiers, we only use 1000 samples for this experiment and only compare MAB-Malware with random action selection. For AV2 and AV3, our approach improves the evasion rate by 8\% to 9\%. For AV1, although the evasion rates are almost the same in the end, the evasion rate of our approach is faster in the beginning. 


From the evasion rates for EMBER and MalConv, we can find out that MAB-Malware can easily mislead pure machine learning models. The reason is that besides finding good actions, MAB-Malware is also good at preserving good content, which largely affects the data distribution of the generated variants. It may indicate that AV2 and AV3 make more use of machine learning models than AV1 in their static detection. We will further infer the root cause of the evasions for different AVs in Section~\ref{sec:explanation}.

\begin{figure*}[t]
\begin{tabular}{ccc}
\includegraphics[valign=T,width=40ex]{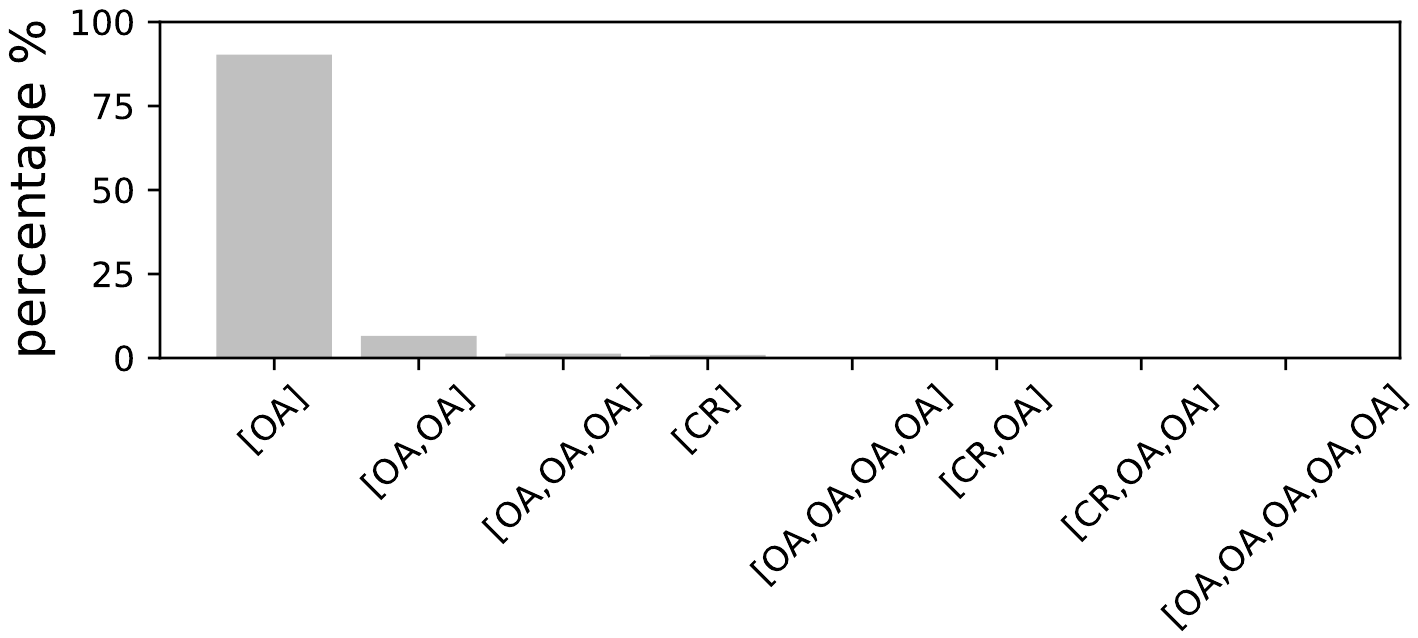} &
\includegraphics[valign=T,width=40ex]{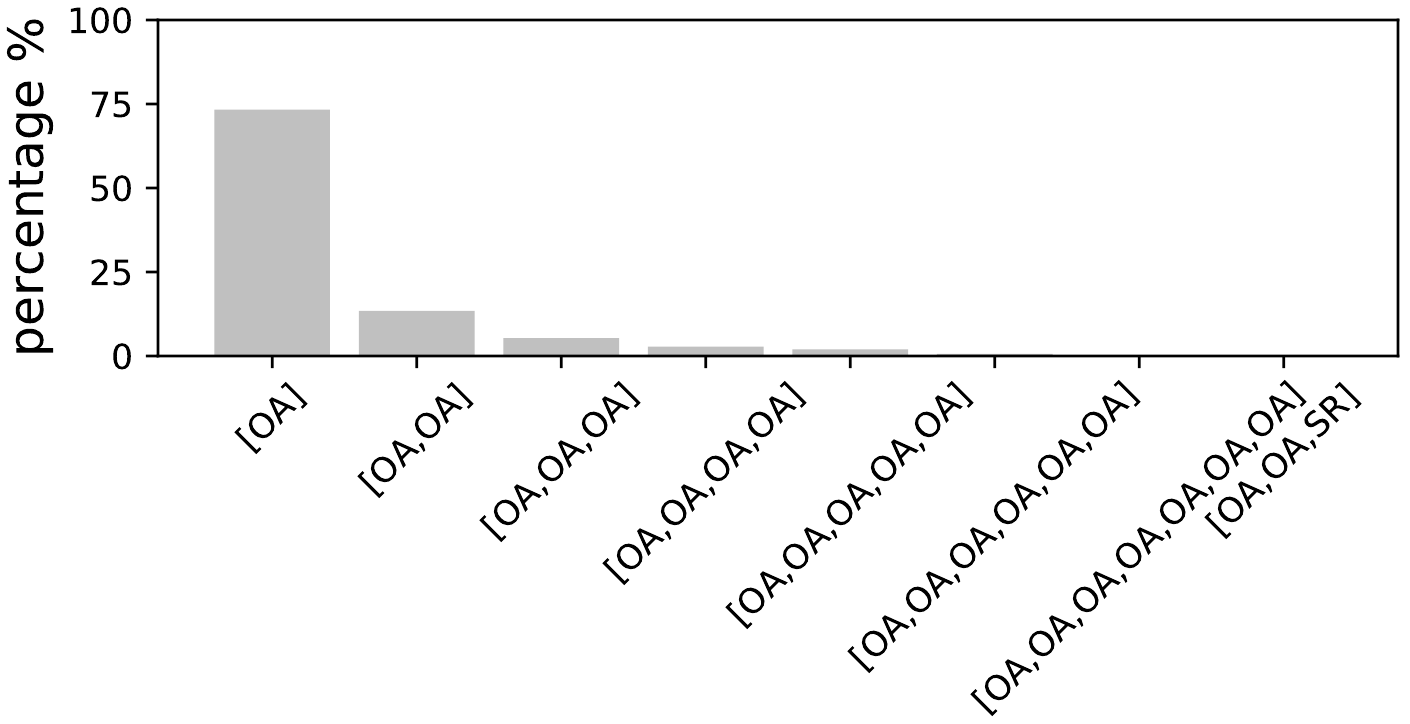} & 
\includegraphics[valign=T,width=40ex]{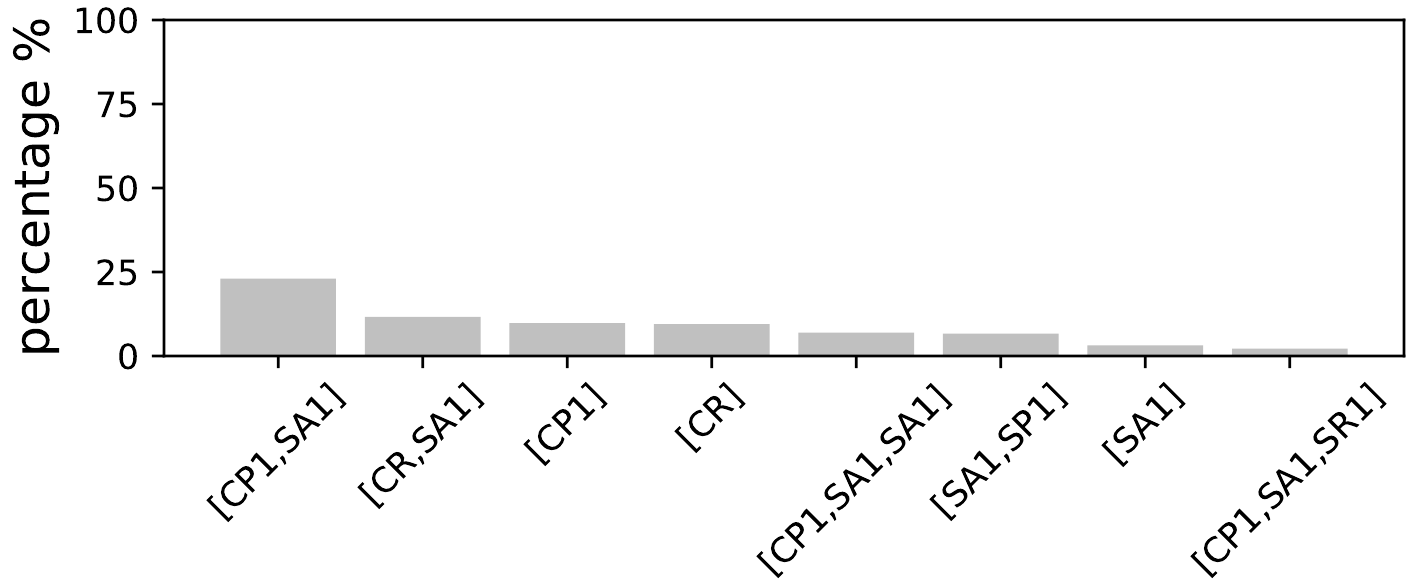} \\
(a) MalConv & (b) EMBER & (c) AV1 \\[6pt]
\includegraphics[valign=T,width=40ex]{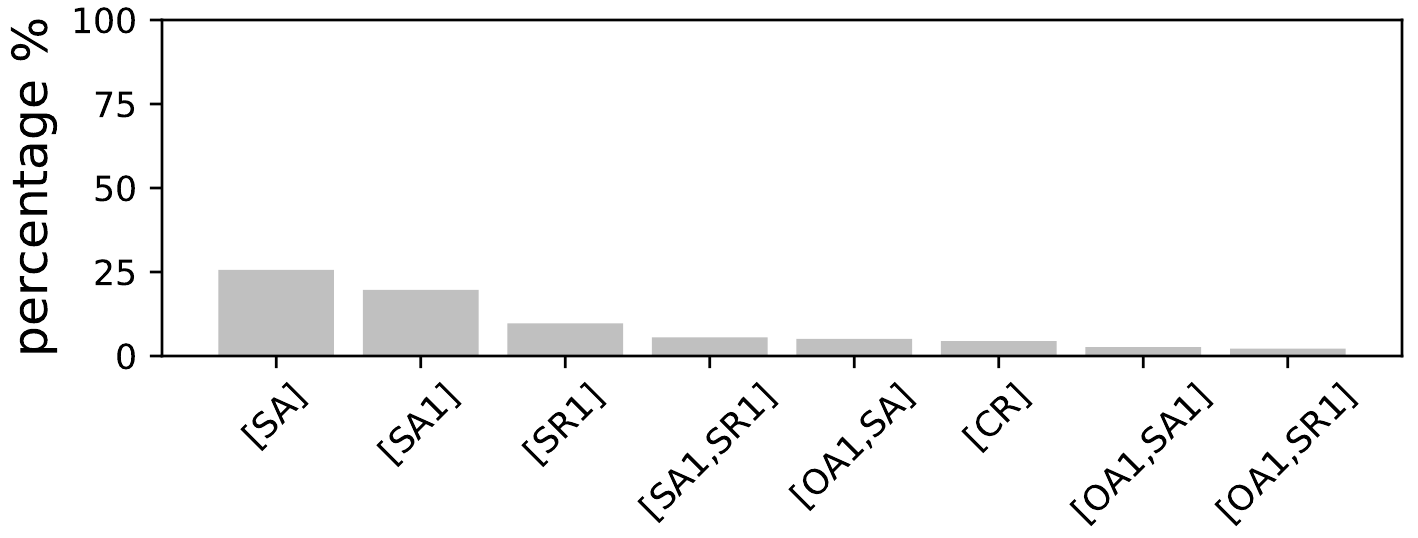} &
\includegraphics[valign=T,width=40ex]{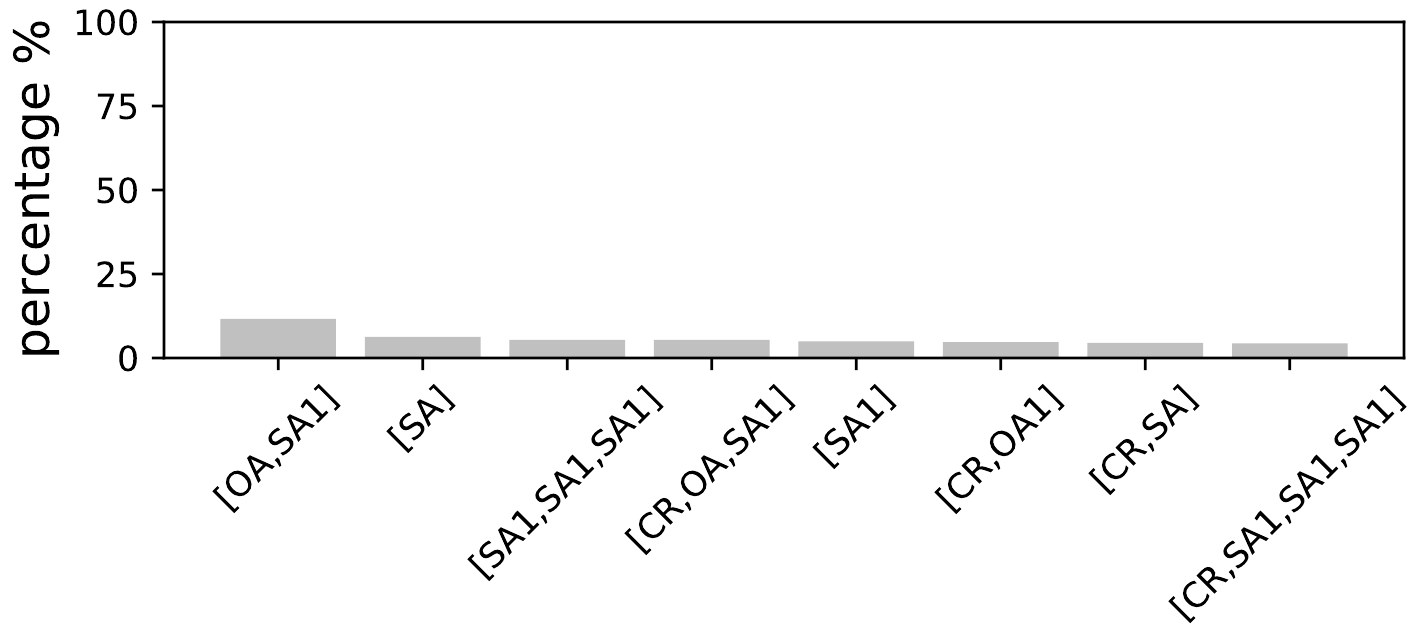} & \\
(d) AV2 & (e) AV3 \\[6pt]
\end{tabular}
\caption{Action Sequences for Adversarial Examples}
\label{fig:action_sequence}
\label{fig:TS_evasion}
\end{figure*}

\paragraph{Number of bytes changed.} The Action Minimizer makes sure that the minimized evasive samples only change minimal content to flip the classification label. So by checking how many bytes we need to change, we can infer the robustness of different malware classifiers. To measure the difference between the minimized evasive example and the original malware, we compute the total number of bytes appended or modified.

\begin{figure*}[ht]
\begin{tabular}{ccc}
\includegraphics[width=35ex]{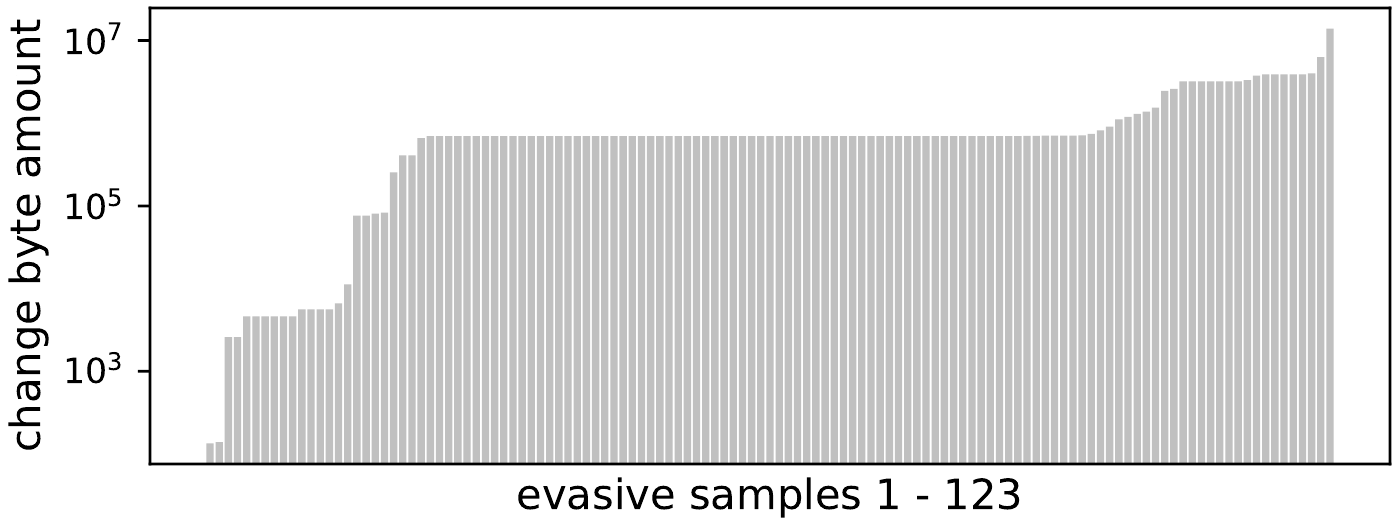}  &
\includegraphics[width=35ex]{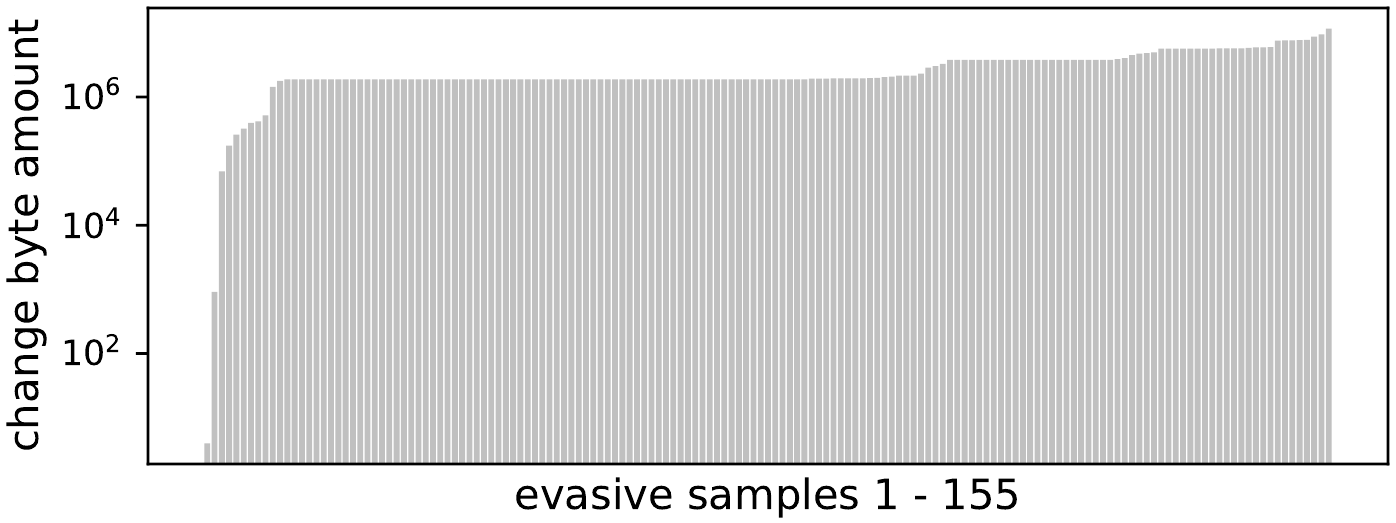} &\\
(a) MalConv & (b) EMBER &\\[6pt]
\includegraphics[width=35ex]{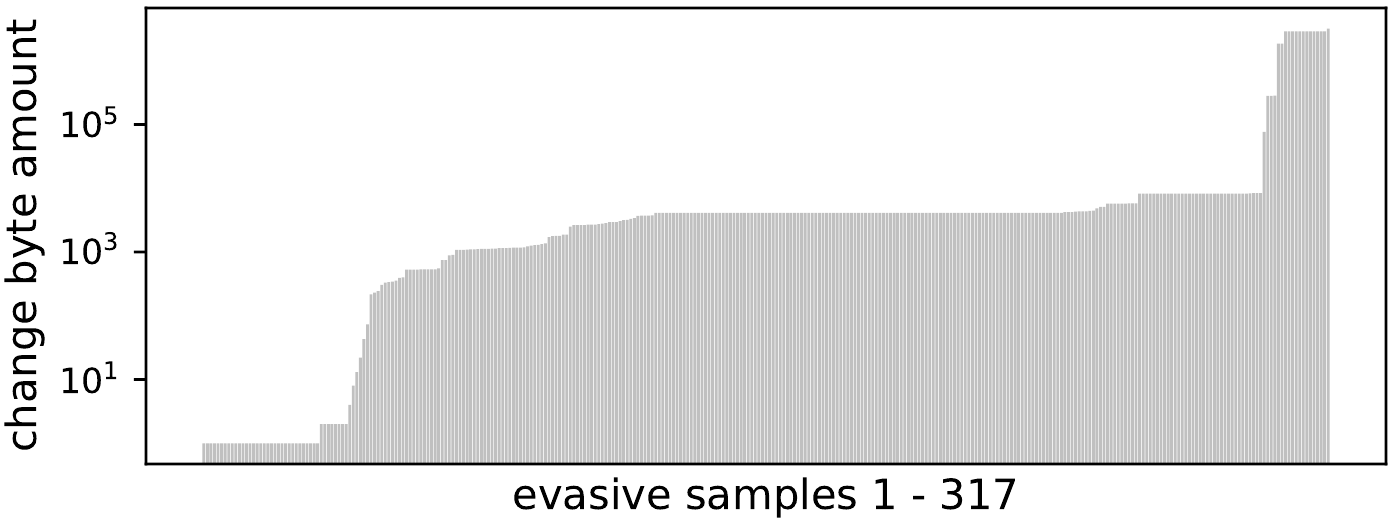} &
\includegraphics[width=35ex]{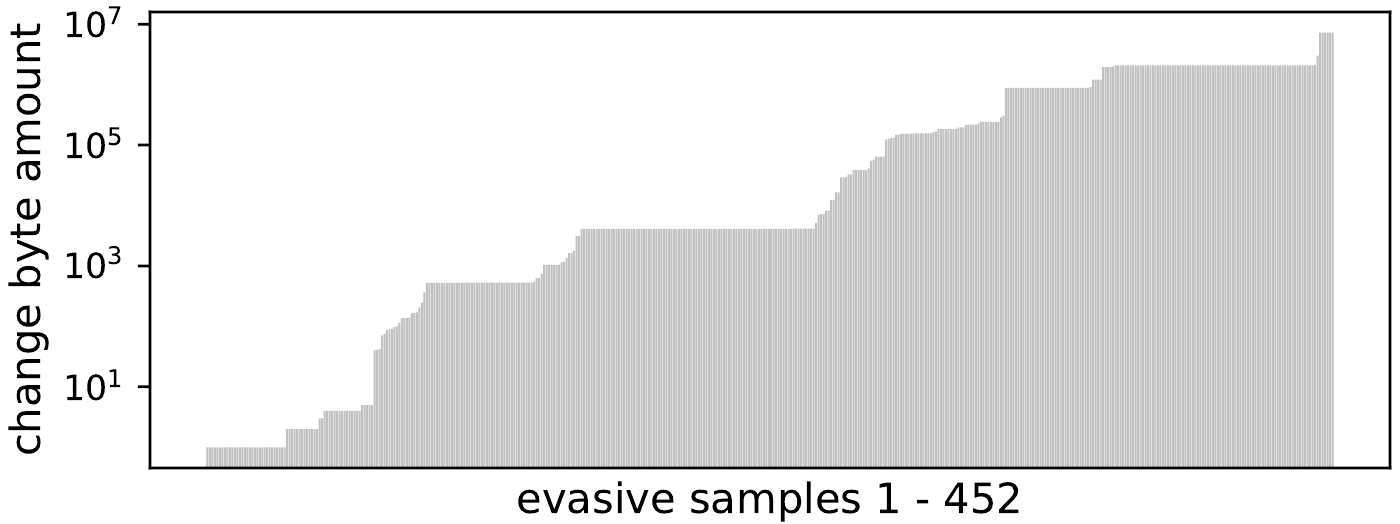} &
\includegraphics[width=35ex]{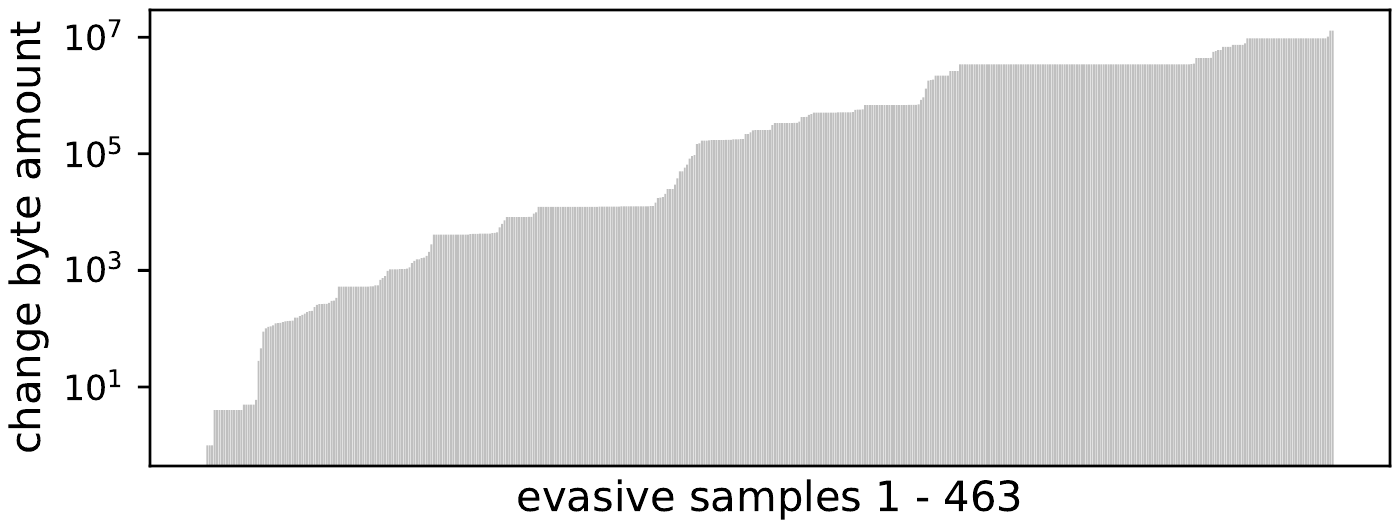}\\
(e) AV1 & (f) AV2 & (G) AV3 \\[6pt]\\
\end{tabular}
\caption{Number of Changed Bytes of Adversarial Examples. By changing only one byte, adversarial examples are created for all tested AVs, with the exception of EMBER, which is not vulnerable to a few bytes of change in the binary.} 
\label{fig:changed_bytes}
\end{figure*}

By positioning the samples in a line sorted by byte changes (Figure~\ref{fig:changed_bytes}), we notice that:
\begin{itemize}
    \item  By only changing one byte of the original malware, we can generate 33 for AV1, 32 for AV2, 3 for AV3.
    
    \item Machine learning models are not vulnerable to small changes. However, it does not mean that ML models are more robust than commercial AVs. From previous evasion rate results, we can see ML models are easier to evade using our framework than commercial AVs.
\end{itemize}

\subsection{Testing Functionality Preservation}

To conduct a blackbox attack on malware classifiers, we need an action set that provides several different transformations that change different features, but do not change the functionality. We found that the action set in Gym-Malware, which is implemented using LIEF~\cite{LIEF} library, is not safe. According to our experiment result in Table~\ref{tab:functional_rate}, more than 60\% of the generated binaries after a single action cannot be executed, or behave differently. To solve this problem, we carefully reimplement most actions using the pefile~\cite{pefile} library to avoid many corner cases that may lead to a broken binary. For example, before adding a new section, we check whether there is sufficient space between the last section header entry and the first section. Similarly, we also detect overlay data at the end of the file before appending new section content right after the last section. Only less than 4\% of rewritten samples after one action behave differently from the original samples. The detailed functional rate comparison is shown in Table~\ref{tab:functional_rate} in the evaluation section.

We implement our own action set $\mathcal{A}$ using the pefile library whereas the Gym-Malware rewrites binaries using the LIEF library. We noticed that rewriting a binary with the LIEF library can cause unnecessary changes to the binary that can sometimes result in broken files, thus destroying the functionality of the original malware samples. To compare our actions with the actions from Gym-Malware, we randomly select 50 malware samples from our dataset, create adversarial samples by applying different actions, analyze all variants in the Cuckoo sandbox, and compare the behavior signatures with the original samples.

\begin{table}[htbp]
\caption{Functionality Preservation Rate of the Actions}
\begin{center}
{\footnotesize
\begin{tabular}{l|r|r}
\hline
\multirow{2}{*}{\textbf{Actions}}    & \multicolumn{2}{c}{\textbf{Functional Rate}} \\ \cline{2-3}
                                    & Gym-Malware Actions          & MAB-Malware Actions   \\ \hline
(OA) Overlay Append     & 45/48 (93.75\%)           & 46/48 (95.83\%)               \\
(SP) Section Append     & 11/47 (23.40\%)           & 42/43 (97.67\%)               \\
(SA) Section Add        & 11/47 (23.40\%)           & 39/42 (92.86\%)               \\
(SR) Section Rename     & 11/47 (23.40\%)           & 42/43 (97.67\%)               \\
(RC) Remove Certificate & 1/3 (33.33\%)             & 3/3 (100.00\%)                \\
(RD) Remove Debug       & 5/13 (38.46\%)            & 13/13 (100.00\%)              \\
(BC) Break Checksum     & 9/48 (18.75\%)            & 32/33 (96.97\%)               \\ \hline
\textbf{Average}        & \textbf{93/253 (36.76\%)} & \textbf{217/225 (96.44\%)}    \\ \hline
\end{tabular}
\label{tab:functional_rate}
}
\end{center}
\end{table}

From Table~\ref{tab:functional_rate} we can see that except for the Overlay Append action, most actions in the Gym-Malware framework cause 63.24\% of the rewritten samples to lose functionality. In contrast, only less than 8\% of the rewritten samples using our actions create broken binaries. One reason that the innocuous actions break the binaries is that these actions can override other sections causing corrupted files (see Appendix figure~\ref{fig:break2} for an example).

\subsection{Explanation}
\label{sec:explanation}

\begin{figure}[t]
    \centering
    \includegraphics[width=1\linewidth]{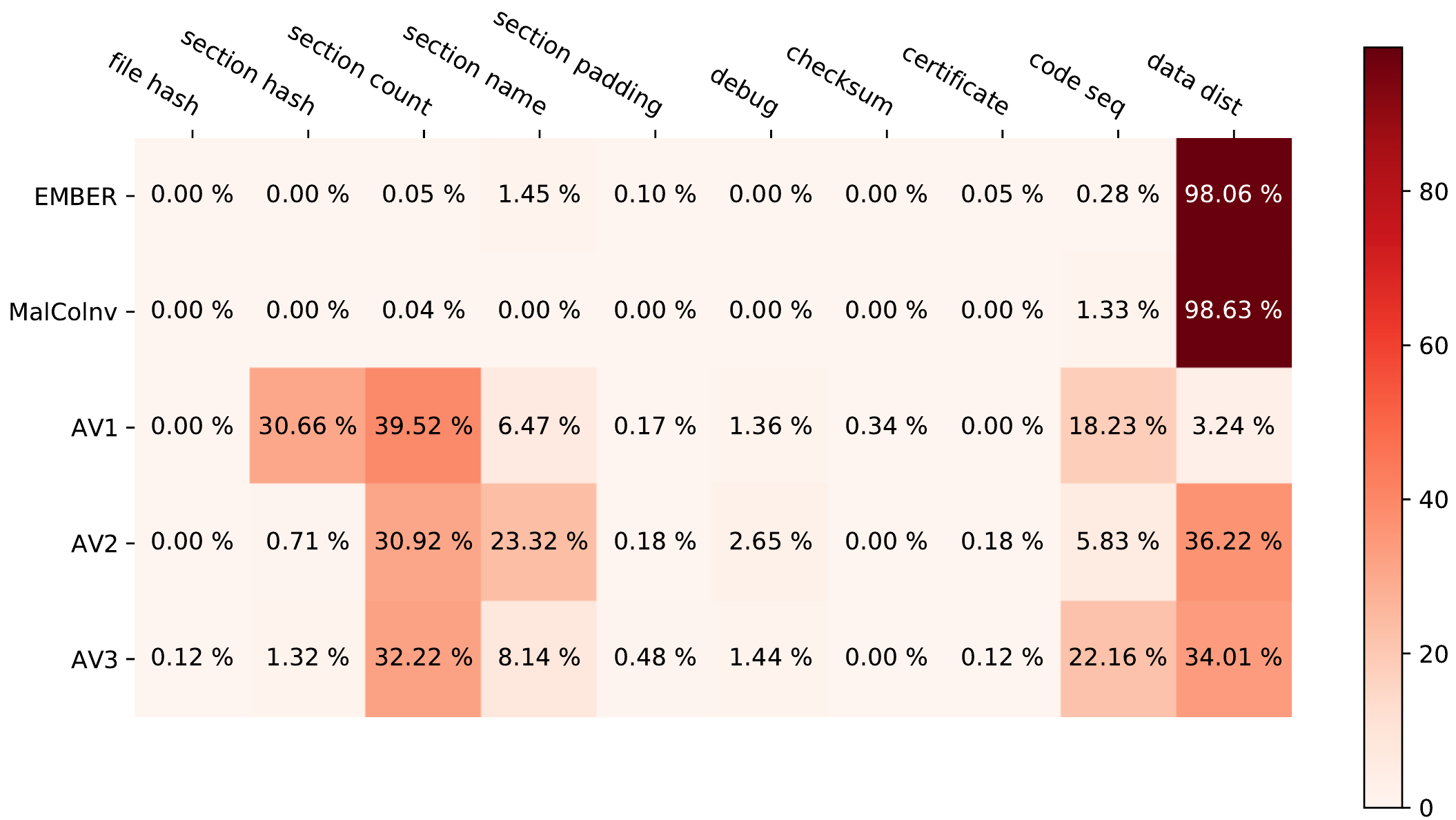}
    \caption{Feature changes that cause evasion.}
    \label{fig:feature_contribution}
\end{figure}

Understanding why evasion happens can help improve the robustness of a classifier against adversarial attacks. For each evasive sample, the Action Minimizer first removes all redundant actions and uses micro-actions to replace the macro-actions. We summarize the most frequent action sequence combination is Figure~\ref{fig:action_sequence}. According to the rules in Figure~\ref{decision_rules}, we can infer the root cause of each evasion, shown in Figure~\ref{fig:feature_contribution}. We found that:


\begin{itemize}
    \item For two machine learning-based classifiers, the most important action is Overlay Append (OA). Other actions that only change a few bytes have almost no effect on them. It shows that the change in data distribution is the root cause of the evasions.
    \item The Section Add 1 Byte (SA1) action plays a significant role in evading all AVs. It indicates that all AVs utilize section count as an important feature for detecting malware.
    \item Comparing to AV2 and AV3, AV1 is also vulnerable to the Code Section Append 1 Byte (CP1) action. CP1 alters the hash of the code section. It indicates AV1 uses code section hash as an important feature for detecting malware.
    \item The Section Rename 1 Byte (SR1) action itself can generate many adversarial examples for AV2. SR1 changes one byte of one section name. It indicates that AV2 relies heavily on the section name for detecting malware.
    \item Comparing with AV2 and AV3, the Section Add (SA) action and the Overlay Append (OA) action have almost no effect on AV1. SA and OA greatly change the data distribution of the original malware samples. It indicates that AV2 and AV3 integrate some machine learning models in static detection. And AV1 mainly uses the signature-based approach to detect malware. 
\end{itemize}

\subsection{Transferability}
\label{sec:transfer}

\begin{figure}[t]
    \centering
    \includegraphics[width=0.8\linewidth]{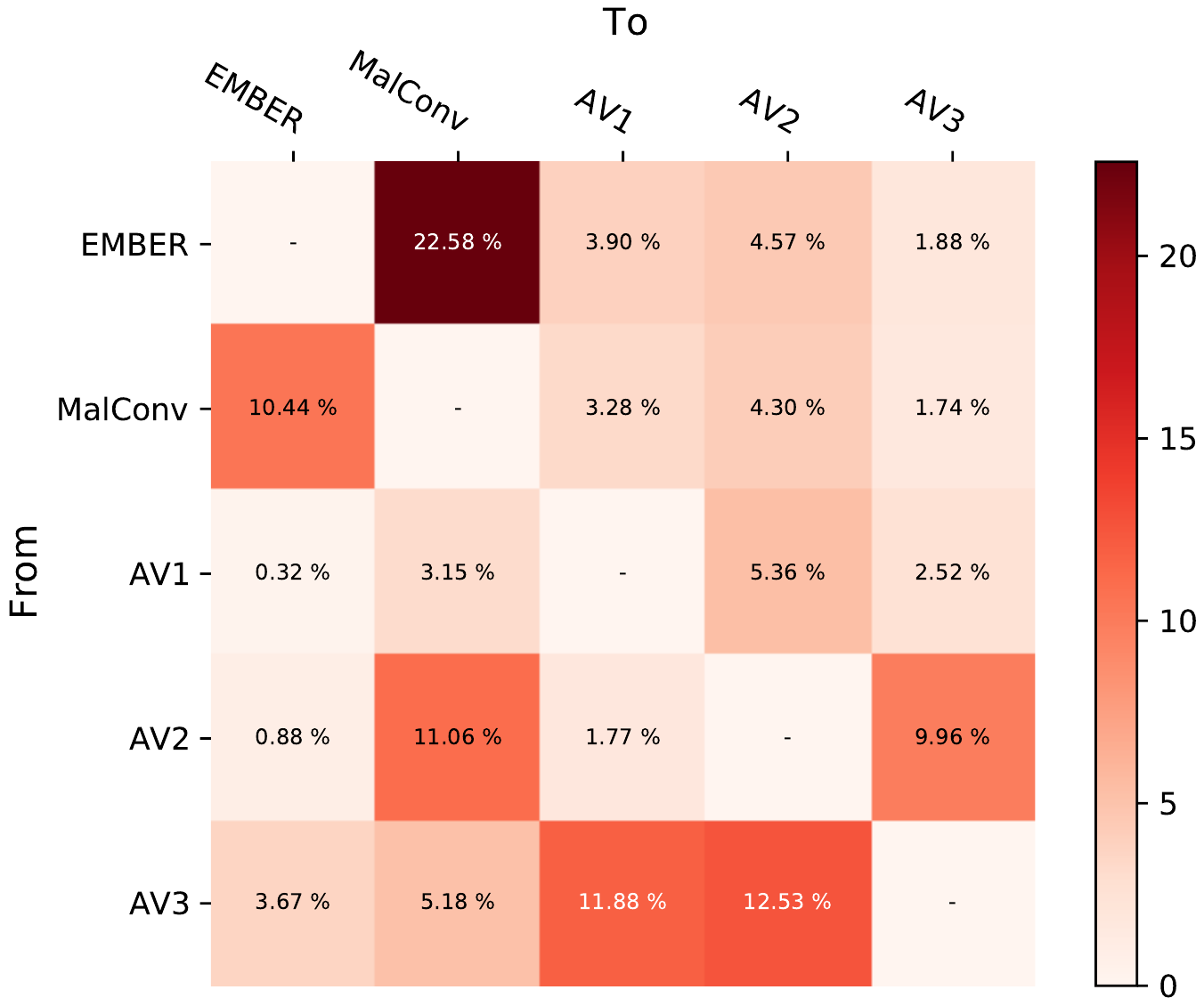}
    \caption{Transferability of Adversarial Samples}
    \label{fig:transfer}
\end{figure}\textbf{}

Transferability refers to the property that allows an adversarial sample that can evade one model can also evade other similar models.  
If the adversarial malware samples are transferable, then evading one malware detector would be enough to evade all malware detectors.

Figure~\ref{fig:transfer} shows the percentage of evasive samples generated for one classifier that can also evade other classifiers. The number in the cell (model A, model B) shows the percentage of evasive samples generated for model A can also evade model B.
We noticed:
\begin{itemize}
    \item The transferability between machine learning models is quite high, although EMBER and MalConv are trained on different features (2350-dimensional extracted features vs. raw bytes), and the architecture is different (decision tree vs. neural network).
    \item Both AV2 and AV3 utilize machine learning models and consider Section count as an important feature. So the transferability between AV2 and AV3 is relatively higher than others.
\end{itemize}
\section{Discussions}
\label{sec:discuss}

\paragraph{Triviality of Defense.} The triviality of the defense depends on the type of attack. To defend the overlay append attack, the defender can ignore the overlay data when training models. To defend against the SA attack, the defender can lower the importance of benign features in models, and only consider malware features. To defend the RD, SR, BC attack, the defender should avoid using such fragile patterns as malware features. However, completely ignoring the trivial features can reduce the accuracy of a malware detector. The code randomization (CR) attack is hard to defend because the defender cannot locate the small snippet of binary that is randomized.

\paragraph{Why do commercial AVs rely on simple features?}
Looking at the result, it seems surprising that trivial changes to malware can evade professionally developed commercial anti-virus systems used by millions of users. Why have not commercial AVs fixed these problems already? One hypothesis could be that adversarial examples are a result of the trade-off between true positive and false-positive rates. The commercial systems need to provide a fast decision while maintaining a low false-positive rate. The relatively simple features, such as file hash or white-listing benign strings can help gain a high accuracy with a low false-positive rate. For example, in their attack against Cylance~\cite{cylance}, researchers noticed that to reduce their false positive rate the Cylance team whitelisted some families of executables, one of them was an online game. So, the whitelisting used to reduce the false-positive rate was enough to create an adversarial attack. Another hypothesis is that anti-virus systems need to protect against the malware of today, instead of focusing on new attacks that are not currently happening. Real adversaries probably use techniques different from the ones used to create ML-based adversarial attacks. A third hypothesis is that anti-virus systems do not rely only on static detection, but also on dynamic and behavioral detection. If all the adversarial malware samples get detected when they are executed, the AVs are not concerned with the static-only adversarial attacks as these attacks cannot infect real users. 

\paragraph{Are adversarial attacks harmful to users?}
We perform a preliminary test to see the extent to which static-only adversarial examples evade the full AV pipeline and infect users.  We create adversarial samples by modifying 30 ransomware samples and test whether the samples that evade static classifiers can infect users’ machines. We hypothesize that the dynamic and behavioral classifiers of the AVs will detect and stop the static adversarial examples when they are executed, thus, posing no real harm to the users. Except for AV2, all the other AVs blocked the execution of adversarial ransomware samples. All of the 30 adversarial ransomware samples evade the behavior detector of AV2; files are encrypted and blackmail messages are shown on the screen. 
However, the online version of AV2 can detect all the samples as AV2 relies heavily on cloud techniques. This represents a potentially new attack surface to investigate in the future, where static-only evasion can sometimes evade the entire AV pipeline and infect users due to the design decision of an AV. 

\paragraph{Recommendation for Antivirus Systems.}
Our attacks demonstrate that static classifiers are easy to evade. So for full protection, commercial AV systems need to rely more on dynamic and behavior-based detection. Even though some papers already demonstrated that dynamic classifiers can be evaded by splitting a malware sample into many different pieces~\cite{dtime,malwash}, currently no one demonstrated a scalable and generalized attack against dynamic classifiers.

\paragraph{Recommendation for Researchers.}
We demonstrate how adversarial examples can be used to explain a complex blackbox system. When training malware classifiers, researchers should use explanation techniques to understand the behavior of the classifiers and check if the learned features are fragile features that can be easily evaded or if they conflict with expert knowledge. We also argue that for security applications, demonstrating harm to real users is crucial to understanding the real ramification of an attack.

\begin{table*}[htbp]
\caption{Adversarial attacks on Windows malware detectors}
\begin{center}
{\small
\begin{tabular}{l|l|l|c|c|c}
\hline
 \textbf{Year}& \textbf{Paper} & \textbf{Target Model} & \textbf{Threat model}    & \textbf{Problem space attack}   & \textbf{Verification} \\ \hline
2019&Ashkenazy et al.~\cite{cylance} & \textbf{AV (Cylance)} & greybox &Yes & No \\
2019 & 
Ceschin et al.~\cite{ceschin2019shallow}    & \textbf{MalConv}, EMBER, \textbf{AVs} & \textbf{blackbox}    & Yes    & \textbf{Yes}    \\
2019& 
Demetrio et al.~\cite{pe5}    & \textbf{MalConv}    & whitebox    & Yes    & No    \\
2019&
Huang et al.~\cite{pe3}    & DNN (API existence)    & grey\&whitebox & \textbf{No}    & No    \\
2019&
Khormali et al.~\cite{pe1}    & CNN (Visualization)    & whitebox    & Yes    & No    \\
2019&
Liu et al.~\cite{pe9}    & CNN (Visualization), RF, SVM    & whitebox    & Yes    & No    \\
2019&
Park et al.~\cite{MalwareObfuscation2019}    & CNN (Visualization), \textbf{MalConv}    & whitebox    & Yes    & No    \\
2019& 
Suciu et al.~\cite{pe7}    & \textbf{MalConv}, \textbf{EMBER}    & whitebox    & Yes    & No    \\
2018&
Al-Dujaili et al.~\cite{pe15}    & DNN    & whitebox    & \textbf{No}    & No    \\
2018& 
Fleshman et al.~\cite{fleshman2018study}    & \textbf{MalConv}, \textbf{AVs}    & \textbf{blackbox}    & Yes    & No    \\
2018&
Hu et al.~\cite{pe18}    & RNN (API call sequence)    & greybox    & \textbf{No}    & No    \\
2018 & 
Kolosnjaji et al.~\cite{pe12}    & \textbf{MalConv}    & whitebox    & Yes    & No    \\
2018 & 
Kreuk et al.~\cite{kreuk2018adversarial}    & \textbf{MalConv}    & whitebox    & Yes    & No    \\
2018&
Rosenberg et al.~\cite{ISRAID2018genericblackbox} & RNN (API call sequence)    & greybox    & Yes    & \textbf{Yes}    \\
2018&
Rosenberg et al.~\cite{pe10}    & RNN (API call sequence)    & greybox    & Yes    & No    \\
2017&
Anderson et al.~\cite{pe14}    & \textbf{EMBER}    & \textbf{blackbox}    & Yes    & No    \\
2017&
Chen et al.~\cite{chen2017adversarialEISI}    & DNN (API existence), \textbf{AVs}    & whitebox    & Yes    & No    \\
2017&
Hu et al.~\cite{hu2017generating}    & DNN (API existence)    & greybox    & \textbf{No}    & No    \\
2017&
Stokes et al.~\cite{stokes2017attack}    & DNN    & whitebox    & \textbf{No}    & No    \\
\hline
\end{tabular}
\label{tab:related_papers}
}
\end{center}
\end{table*}

\paragraph{The generality of our evasive techniques.}
First, our proposed framework conducts a blackbox attack against classifiers. Unlike whitebox attacks, it does not require knowledge of the architecture and parameters of the target classifier. Theoretically, our approach can be used on any malware classifier, as long as the classifier returns a label for testing samples. Second, we attacked 7 representative malware detectors of diverse techniques, including a decision tree-based model (EMBER), a deep learning model (MalConv), and 3 commercial AV engines from top-level AV companies. The significant evasion rate improvement of these detectors proves the generality of our method.

\paragraph{Mitigation using Dynamic Detection}
Our solution cannot bypass dynamic detectors, but we argue that dynamic evasion is another research topic. Static evasion itself is an important research direction because it provides a defense before users execute potentially dangerous programs. This is why ML-based static classifiers, such as EMBER and MalConv, increasingly attract attention in the security community.
\section{Related Work}
\label{sec:related}
Adversarial attacks on machine learning is a rapidly growing field. Since 2014, there has been more than 1400 papers on adversarial attacks and defense.\footnote{\url{https://nicholas.carlini.com/writing/2019/all-adversarial-example-papers.html}} However, only about 42 papers focused on the malware domain, the rest focuses on the image domain. These works performed attacks on Android malware~\cite{android1,android2,android3,android4,android5,android6,android7,papernot2017practical,android8,pierazzi2019intriguing,fass2019hidenoseek,yang2017malware,demontis2019adversarial}, PDF malware~\cite{chen2017adversarialEISI,ehteshamifar2019easy,xu2016automatically,maiorca2018towards}, Windows malware (PE files)~\cite{pe1,pe2,pe3,pe4,pe5,pe6,pe7,pe9,pe10,pe11,pe12,pe13,pe14,pe15,pe16,pe17,pe18,pe19}, IoT malware~\cite{iot} and Flash-based malware~\cite{flash}. 
We compare ourselves with the papers on attacking Windows malware detectors (See Table~\ref{tab:related_papers}). In this domain, only three papers performed blackbox attacks~\cite{ceschin2019shallow,fleshman2018study,pe14}  where an adversary has only external access to the malware detector and also tested their approach against commercial AVs. 
Ceschin et al.~\cite{ceschin2019shallow} developed a white-box attack using open-source models and then submitted the examples to VirusTotal. We refrained from submitting adversarial samples to VirusTotal because many AVs use these samples to retrain their model and the adversarial samples can cause model poisoning. Fleshman et al.~\cite{fleshman2018study} performed blackbox random attacks against four commercial AVs. Anderson et al.~\cite{pe14} propose a reinforcement learning framework gym-malware to performed blackbox attacks against malware classifier EMBER. However, their method only shows about 15\% improvement over random selection. Our attack demonstrates a much higher evasion rate against commercial AVs.  We also perform an in-depth analysis of the AVs to understand why evasion was successful.
Ashkenazy et al.~\cite{cylance} performed a targeted attack against Cylance, a machine learning-based malware detector, which cannot be generalized to other AVs. 

Only two papers verified the functionalities of the adversarial malware samples using the Cuckoo sandbox~\cite{ceschin2019shallow,ISRAID2018genericblackbox}. However, none of them used adversarial examples to interpret how anti-virus systems work.  Our work is the first to generate minimized adversarial examples that can be used for blackbox interpretation of anti-virus systems. 
\section{Conclusion}
\label{sec:conclude}
We design a reinforcement learning guided framework MAB-Malware to perform adversarial attacks on state-of-the-art machine learning models for malware classification and top commercial antivirus static classifiers. We model the action selection problem as a multi-armed bandit problem. During the attack, MAB-Malware infers the property of actions and dynamically adding new machines with unseen successful content. It finds an optimal balance between exploitation and exploration to maximize the evasion rate within limited trials. The Action Minimization module of MAB-Malware filters out the actions that are ineffective for adversarial sample generation and only change minimal features, so our framework can also be used to explain why evasion occurs. For each commercial antivirus system, we compute the effectiveness of each action and the key features that cause evasions. Our results show that MAB-Malware largely improves the evasion rate over other reinforcement learning frameworks and that some of the adversarial attacks are transferable between different antivirus systems that are similar to one another.


\bibliographystyle{ACM-Reference-Format}
\bibliography{reference}


\begin{thebibliography}{69}


\ifx \showCODEN    \undefined \def \showCODEN     #1{\unskip}     \fi
\ifx \showDOI      \undefined \def \showDOI       #1{#1}\fi
\ifx \showISBNx    \undefined \def \showISBNx     #1{\unskip}     \fi
\ifx \showISBNxiii \undefined \def \showISBNxiii  #1{\unskip}     \fi
\ifx \showISSN     \undefined \def \showISSN      #1{\unskip}     \fi
\ifx \showLCCN     \undefined \def \showLCCN      #1{\unskip}     \fi
\ifx \shownote     \undefined \def \shownote      #1{#1}          \fi
\ifx \showarticletitle \undefined \def \showarticletitle #1{#1}   \fi
\ifx \showURL      \undefined \def \showURL       {\relax}        \fi
\providecommand\bibfield[2]{#2}
\providecommand\bibinfo[2]{#2}
\providecommand\natexlab[1]{#1}
\providecommand\showeprint[2][]{arXiv:#2}

\bibitem[\protect\citeauthoryear{??}{bet}{[n.d.]}]%
        {beta}
 \bibinfo{year}{[n.d.]}\natexlab{}.
\newblock \bibinfo{title}{{Beta distribution}}.
\newblock
  \bibinfo{howpublished}{\url{https://en.wikipedia.org/wiki/Beta\_distribution}}.
\newblock


\bibitem[\protect\citeauthoryear{Abusnaina, Khormali, Alasmary, Park, Anwar,
  Meteriz, and Mohaisen}{Abusnaina et~al\mbox{.}}{2019}]%
        {iot}
\bibfield{author}{\bibinfo{person}{Ahmed Abusnaina}, \bibinfo{person}{Aminollah
  Khormali}, \bibinfo{person}{Hisham Alasmary}, \bibinfo{person}{Jeman Park},
  \bibinfo{person}{Afsah Anwar}, \bibinfo{person}{Ulku Meteriz}, {and}
  \bibinfo{person}{Aziz Mohaisen}.} \bibinfo{year}{2019}\natexlab{}.
\newblock \showarticletitle{Examining Adversarial Learning against Graph-based
  IoT Malware Detection Systems}.
\newblock \bibinfo{journal}{\emph{arXiv preprint arXiv:1902.04416}}
  (\bibinfo{year}{2019}).
\newblock


\bibitem[\protect\citeauthoryear{Al-Dujaili, Huang, Hemberg, and
  O’Reilly}{Al-Dujaili et~al\mbox{.}}{2018}]%
        {pe15}
\bibfield{author}{\bibinfo{person}{Abdullah Al-Dujaili}, \bibinfo{person}{Alex
  Huang}, \bibinfo{person}{Erik Hemberg}, {and} \bibinfo{person}{Una-May
  O’Reilly}.} \bibinfo{year}{2018}\natexlab{}.
\newblock \showarticletitle{Adversarial deep learning for robust detection of
  binary encoded malware}. In \bibinfo{booktitle}{\emph{2018 IEEE Security and
  Privacy Workshops (SPW)}}. IEEE, \bibinfo{pages}{76--82}.
\newblock


\bibitem[\protect\citeauthoryear{Anderson, Kharkar, Filar, Evans, and
  Roth}{Anderson et~al\mbox{.}}{2018}]%
        {pe14}
\bibfield{author}{\bibinfo{person}{Hyrum~S Anderson}, \bibinfo{person}{Anant
  Kharkar}, \bibinfo{person}{Bobby Filar}, \bibinfo{person}{David Evans}, {and}
  \bibinfo{person}{Phil Roth}.} \bibinfo{year}{2018}\natexlab{}.
\newblock \showarticletitle{Learning to evade static PE machine learning
  malware models via reinforcement learning}.
\newblock \bibinfo{journal}{\emph{arXiv preprint arXiv:1801.08917}}
  (\bibinfo{year}{2018}).
\newblock


\bibitem[\protect\citeauthoryear{Anderson and Roth}{Anderson and Roth}{2018}]%
        {ember}
\bibfield{author}{\bibinfo{person}{Hyrum~S Anderson} {and}
  \bibinfo{person}{Phil Roth}.} \bibinfo{year}{2018}\natexlab{}.
\newblock \showarticletitle{Ember: an open dataset for training static PE
  malware machine learning models}.
\newblock \bibinfo{journal}{\emph{arXiv preprint arXiv:1804.04637}}
  (\bibinfo{year}{2018}).
\newblock


\bibitem[\protect\citeauthoryear{Ashkenazy and Zini}{Ashkenazy and
  Zini}{2019}]%
        {cylance}
\bibfield{author}{\bibinfo{person}{Adi Ashkenazy} {and} \bibinfo{person}{Shahar
  Zini}.} \bibinfo{year}{2019}\natexlab{}.
\newblock \bibinfo{title}{Cylance, I Kill You!}
\newblock
  \bibinfo{howpublished}{\url{https://skylightcyber.com/2019/07/18/cylance-i-kill-you/}}.
\newblock


\bibitem[\protect\citeauthoryear{Avast}{Avast}{2018}]%
        {avast_link}
Avast \bibinfo{year}{2018}\natexlab{}.
\newblock \bibinfo{title}{AI \& Machine Learning}.
\newblock
  \bibinfo{howpublished}{\url{https://www.avast.com/en-us/technology/ai-and-machine-learning}}.
\newblock


\bibitem[\protect\citeauthoryear{Carlini, Athalye, Papernot, Brendel, Rauber,
  Tsipras, Goodfellow, and Madry}{Carlini et~al\mbox{.}}{2019}]%
        {carlini2019evaluating}
\bibfield{author}{\bibinfo{person}{Nicholas Carlini}, \bibinfo{person}{Anish
  Athalye}, \bibinfo{person}{Nicolas Papernot}, \bibinfo{person}{Wieland
  Brendel}, \bibinfo{person}{Jonas Rauber}, \bibinfo{person}{Dimitris Tsipras},
  \bibinfo{person}{Ian Goodfellow}, {and} \bibinfo{person}{Aleksander Madry}.}
  \bibinfo{year}{2019}\natexlab{}.
\newblock \showarticletitle{On evaluating adversarial robustness}.
\newblock \bibinfo{journal}{\emph{arXiv preprint arXiv:1902.06705}}
  (\bibinfo{year}{2019}).
\newblock


\bibitem[\protect\citeauthoryear{Carrera}{Carrera}{2016}]%
        {pefile}
\bibfield{author}{\bibinfo{person}{Ero Carrera}.}
  \bibinfo{year}{2016}\natexlab{}.
\newblock \bibinfo{title}{{pefile}}.
\newblock \bibinfo{howpublished}{\url{https://github.com/erocarrera/pefile}}.
\newblock


\bibitem[\protect\citeauthoryear{Castro, Schmitt, and Dreo}{Castro
  et~al\mbox{.}}{2019}]%
        {castro2019aimed}
\bibfield{author}{\bibinfo{person}{Raphael~Labaca Castro},
  \bibinfo{person}{Corinna Schmitt}, {and} \bibinfo{person}{Gabi Dreo}.}
  \bibinfo{year}{2019}\natexlab{}.
\newblock \showarticletitle{AIMED: Evolving Malware with Genetic Programming to
  Evade Detection}. In \bibinfo{booktitle}{\emph{2019 18th IEEE International
  Conference On Trust, Security And Privacy In Computing And
  Communications/13th IEEE International Conference On Big Data Science And
  Engineering (TrustCom/BigDataSE)}}. IEEE, \bibinfo{pages}{240--247}.
\newblock


\bibitem[\protect\citeauthoryear{Ceschin, Botacin, Gomes, Oliveira, and
  Gr{\'e}gio}{Ceschin et~al\mbox{.}}{2019}]%
        {ceschin2019shallow}
\bibfield{author}{\bibinfo{person}{Fabr{\'\i}cio Ceschin},
  \bibinfo{person}{Marcus Botacin}, \bibinfo{person}{Heitor~Murilo Gomes},
  \bibinfo{person}{Luiz~S Oliveira}, {and} \bibinfo{person}{Andr{\'e}
  Gr{\'e}gio}.} \bibinfo{year}{2019}\natexlab{}.
\newblock \showarticletitle{Shallow Security: on the Creation of Adversarial
  Variants to Evade Machine Learning-Based Malware Detectors}. In
  \bibinfo{booktitle}{\emph{Proceedings of the 3rd Reversing and
  Offensive-oriented Trends Symposium}}. \bibinfo{pages}{1--9}.
\newblock


\bibitem[\protect\citeauthoryear{Chapelle and Li}{Chapelle and Li}{2011}]%
        {chapelle2011empirical}
\bibfield{author}{\bibinfo{person}{Olivier Chapelle} {and}
  \bibinfo{person}{Lihong Li}.} \bibinfo{year}{2011}\natexlab{}.
\newblock \showarticletitle{An empirical evaluation of thompson sampling}. In
  \bibinfo{booktitle}{\emph{Advances in neural information processing
  systems}}. \bibinfo{pages}{2249--2257}.
\newblock


\bibitem[\protect\citeauthoryear{Chen}{Chen}{2019}]%
        {pe2}
\bibfield{author}{\bibinfo{person}{Li Chen}.} \bibinfo{year}{2019}\natexlab{}.
\newblock \showarticletitle{Understanding the efficacy, reliability and
  resiliency of computer vision techniques for malware detection and future
  research directions}.
\newblock \bibinfo{journal}{\emph{arXiv preprint arXiv:1904.10504}}
  (\bibinfo{year}{2019}).
\newblock


\bibitem[\protect\citeauthoryear{Chen, Ye, and Bourlai}{Chen
  et~al\mbox{.}}{2017}]%
        {chen2017adversarialEISI}
\bibfield{author}{\bibinfo{person}{Lingwei Chen}, \bibinfo{person}{Yanfang Ye},
  {and} \bibinfo{person}{Thirimachos Bourlai}.}
  \bibinfo{year}{2017}\natexlab{}.
\newblock \showarticletitle{Adversarial machine learning in malware detection:
  Arms race between evasion attack and defense}. In
  \bibinfo{booktitle}{\emph{2017 European Intelligence and Security Informatics
  Conference (EISIC)}}. IEEE, \bibinfo{pages}{99--106}.
\newblock


\bibitem[\protect\citeauthoryear{Dahl, Stokes, Deng, and Yu}{Dahl
  et~al\mbox{.}}{2013}]%
        {dahl2013large}
\bibfield{author}{\bibinfo{person}{George~E Dahl}, \bibinfo{person}{Jack~W
  Stokes}, \bibinfo{person}{Li Deng}, {and} \bibinfo{person}{Dong Yu}.}
  \bibinfo{year}{2013}\natexlab{}.
\newblock \showarticletitle{Large-scale malware classification using random
  projections and neural networks}. In \bibinfo{booktitle}{\emph{2013 IEEE
  International Conference on Acoustics, Speech and Signal Processing}}. IEEE,
  \bibinfo{pages}{3422--3426}.
\newblock


\bibitem[\protect\citeauthoryear{Daniel, Haidar, and Bülent}{Daniel
  et~al\mbox{.}}{2019}]%
        {MalwareObfuscation2019}
\bibfield{author}{\bibinfo{person}{Park Daniel}, \bibinfo{person}{Khan Haidar},
  {and} \bibinfo{person}{Yener Bülent}.} \bibinfo{year}{2019}\natexlab{}.
\newblock \showarticletitle{Generation \& Evaluation of Adversarial Examples
  for Malware Obfuscation}.
\newblock \bibinfo{journal}{\emph{arXiv preprint arXiv:1904.04802}}
  (\bibinfo{year}{2019}).
\newblock


\bibitem[\protect\citeauthoryear{Demetrio, Biggio, Lagorio, Roli, and
  Armando}{Demetrio et~al\mbox{.}}{2019}]%
        {pe5}
\bibfield{author}{\bibinfo{person}{Luca Demetrio}, \bibinfo{person}{Battista
  Biggio}, \bibinfo{person}{Giovanni Lagorio}, \bibinfo{person}{Fabio Roli},
  {and} \bibinfo{person}{Alessandro Armando}.} \bibinfo{year}{2019}\natexlab{}.
\newblock \showarticletitle{Explaining Vulnerabilities of Deep Learning to
  Adversarial Malware Binaries}.
\newblock \bibinfo{journal}{\emph{arXiv preprint arXiv:1901.03583}}
  (\bibinfo{year}{2019}).
\newblock


\bibitem[\protect\citeauthoryear{Demetrio, Biggio, Lagorio, Roli, and
  Armando}{Demetrio et~al\mbox{.}}{2020}]%
        {Demetrio2020FunctionalitypreservingBO}
\bibfield{author}{\bibinfo{person}{Luca Demetrio}, \bibinfo{person}{B. Biggio},
  \bibinfo{person}{Giovanni Lagorio}, \bibinfo{person}{F. Roli}, {and}
  \bibinfo{person}{A. Armando}.} \bibinfo{year}{2020}\natexlab{}.
\newblock \showarticletitle{Functionality-preserving Black-box Optimization of
  Adversarial Windows Malware.}
\newblock \bibinfo{journal}{\emph{arXiv: Cryptography and Security}}
  (\bibinfo{year}{2020}).
\newblock


\bibitem[\protect\citeauthoryear{Demontis, Melis, Biggio, Maiorca, Arp, Rieck,
  Corona, Giacinto, and Roli}{Demontis et~al\mbox{.}}{2017}]%
        {android6}
\bibfield{author}{\bibinfo{person}{Ambra Demontis}, \bibinfo{person}{Marco
  Melis}, \bibinfo{person}{Battista Biggio}, \bibinfo{person}{Davide Maiorca},
  \bibinfo{person}{Daniel Arp}, \bibinfo{person}{Konrad Rieck},
  \bibinfo{person}{Igino Corona}, \bibinfo{person}{Giorgio Giacinto}, {and}
  \bibinfo{person}{Fabio Roli}.} \bibinfo{year}{2017}\natexlab{}.
\newblock \showarticletitle{Yes, machine learning can be more secure! a case
  study on android malware detection}.
\newblock \bibinfo{journal}{\emph{IEEE Transactions on Dependable and Secure
  Computing}} (\bibinfo{year}{2017}).
\newblock


\bibitem[\protect\citeauthoryear{Demontis, Melis, Pintor, Jagielski, Biggio,
  Oprea, Nita-Rotaru, and Roli}{Demontis et~al\mbox{.}}{2019}]%
        {demontis2019adversarial}
\bibfield{author}{\bibinfo{person}{Ambra Demontis}, \bibinfo{person}{Marco
  Melis}, \bibinfo{person}{Maura Pintor}, \bibinfo{person}{Matthew Jagielski},
  \bibinfo{person}{Battista Biggio}, \bibinfo{person}{Alina Oprea},
  \bibinfo{person}{Cristina Nita-Rotaru}, {and} \bibinfo{person}{Fabio Roli}.}
  \bibinfo{year}{2019}\natexlab{}.
\newblock \showarticletitle{Why do adversarial attacks transfer? explaining
  transferability of evasion and poisoning attacks}. In
  \bibinfo{booktitle}{\emph{28th $\{$USENIX$\}$ Security Symposium
  ($\{$USENIX$\}$ Security 19)}}. \bibinfo{pages}{321--338}.
\newblock


\bibitem[\protect\citeauthoryear{Ehteshamifar, Barresi, Gross, and
  Pradel}{Ehteshamifar et~al\mbox{.}}{2019}]%
        {ehteshamifar2019easy}
\bibfield{author}{\bibinfo{person}{Saeed Ehteshamifar},
  \bibinfo{person}{Antonio Barresi}, \bibinfo{person}{Thomas~R Gross}, {and}
  \bibinfo{person}{Michael Pradel}.} \bibinfo{year}{2019}\natexlab{}.
\newblock \showarticletitle{Easy to Fool? Testing the Anti-evasion Capabilities
  of PDF Malware Scanners}.
\newblock \bibinfo{journal}{\emph{arXiv preprint arXiv:1901.05674}}
  (\bibinfo{year}{2019}).
\newblock


\bibitem[\protect\citeauthoryear{Fass, Backes, and Stock}{Fass
  et~al\mbox{.}}{2019}]%
        {fass2019hidenoseek}
\bibfield{author}{\bibinfo{person}{Aurore Fass}, \bibinfo{person}{Michael
  Backes}, {and} \bibinfo{person}{Ben Stock}.} \bibinfo{year}{2019}\natexlab{}.
\newblock \showarticletitle{Hidenoseek: Camouflaging malicious javascript in
  benign asts}. In \bibinfo{booktitle}{\emph{Proceedings of the 2019 ACM SIGSAC
  Conference on Computer and Communications Security}}.
  \bibinfo{pages}{1899--1913}.
\newblock


\bibitem[\protect\citeauthoryear{Fleshman, Raff, Zak, McLean, and
  Nicholas}{Fleshman et~al\mbox{.}}{2018}]%
        {fleshman2018study}
\bibfield{author}{\bibinfo{person}{William Fleshman}, \bibinfo{person}{Edward
  Raff}, \bibinfo{person}{Richard Zak}, \bibinfo{person}{Mark McLean}, {and}
  \bibinfo{person}{Charles Nicholas}.} \bibinfo{year}{2018}\natexlab{}.
\newblock \showarticletitle{Static malware detection \& subterfuge: Quantifying
  the robustness of machine learning and current anti-virus}. In
  \bibinfo{booktitle}{\emph{2018 13th International Conference on Malicious and
  Unwanted Software (MALWARE)}}. IEEE, \bibinfo{pages}{1--10}.
\newblock


\bibitem[\protect\citeauthoryear{Hu and Tan}{Hu and Tan}{2017a}]%
        {pe19}
\bibfield{author}{\bibinfo{person}{Weiwei Hu} {and} \bibinfo{person}{Ying
  Tan}.} \bibinfo{year}{2017}\natexlab{a}.
\newblock \showarticletitle{Generating adversarial malware examples for
  black-box attacks based on GAN}.
\newblock \bibinfo{journal}{\emph{arXiv preprint arXiv:1702.05983}}
  (\bibinfo{year}{2017}).
\newblock


\bibitem[\protect\citeauthoryear{Hu and Tan}{Hu and Tan}{2017b}]%
        {hu2017generating}
\bibfield{author}{\bibinfo{person}{Weiwei Hu} {and} \bibinfo{person}{Ying
  Tan}.} \bibinfo{year}{2017}\natexlab{b}.
\newblock \showarticletitle{Generating adversarial malware examples for
  black-box attacks based on GAN}.
\newblock \bibinfo{journal}{\emph{arXiv preprint arXiv:1702.05983}}
  (\bibinfo{year}{2017}).
\newblock


\bibitem[\protect\citeauthoryear{Hu and Tan}{Hu and Tan}{2018}]%
        {pe18}
\bibfield{author}{\bibinfo{person}{Weiwei Hu} {and} \bibinfo{person}{Ying
  Tan}.} \bibinfo{year}{2018}\natexlab{}.
\newblock \showarticletitle{Black-box attacks against RNN based malware
  detection algorithms}. In \bibinfo{booktitle}{\emph{Workshops at the
  Thirty-Second AAAI Conference on Artificial Intelligence}}.
\newblock


\bibitem[\protect\citeauthoryear{Huang, Al-Dujaili, Hemberg, and
  O'Reilly}{Huang et~al\mbox{.}}{2018}]%
        {pe11}
\bibfield{author}{\bibinfo{person}{Alex Huang}, \bibinfo{person}{Abdullah
  Al-Dujaili}, \bibinfo{person}{Erik Hemberg}, {and} \bibinfo{person}{Una-May
  O'Reilly}.} \bibinfo{year}{2018}\natexlab{}.
\newblock \showarticletitle{On visual hallmarks of robustness to adversarial
  malware}.
\newblock \bibinfo{journal}{\emph{arXiv preprint arXiv:1805.03553}}
  (\bibinfo{year}{2018}).
\newblock


\bibitem[\protect\citeauthoryear{Huang, Verma, Fralick, Infantec-Lopez, Kumar,
  and Woodward}{Huang et~al\mbox{.}}{2019}]%
        {pe3}
\bibfield{author}{\bibinfo{person}{Yonghong Huang}, \bibinfo{person}{Utkarsh
  Verma}, \bibinfo{person}{Celeste Fralick}, \bibinfo{person}{Gabriel
  Infantec-Lopez}, \bibinfo{person}{Brajesh Kumar}, {and} \bibinfo{person}{Carl
  Woodward}.} \bibinfo{year}{2019}\natexlab{}.
\newblock \showarticletitle{Malware Evasion Attack and Defense}.
\newblock \bibinfo{journal}{\emph{2019 49th Annual IEEE/IFIP International
  Conference on Dependable Systems and Networks Workshops (DSN-W)}}
  (\bibinfo{date}{Jun} \bibinfo{year}{2019}).
\newblock
\showISBNx{9781728130309}
\urldef\tempurl%
\url{https://doi.org/10.1109/dsn-w.2019.00014}
\showDOI{\tempurl}


\bibitem[\protect\citeauthoryear{Ispoglou and Payer}{Ispoglou and
  Payer}{2016}]%
        {malwash}
\bibfield{author}{\bibinfo{person}{Kyriakos~K. Ispoglou} {and}
  \bibinfo{person}{Mathias Payer}.} \bibinfo{year}{2016}\natexlab{}.
\newblock \showarticletitle{malWASH: Washing Malware to Evade Dynamic
  Analysis}. In \bibinfo{booktitle}{\emph{10th {USENIX} Workshop on Offensive
  Technologies ({WOOT} 16)}}. \bibinfo{publisher}{{USENIX} Association},
  \bibinfo{address}{Austin, TX}.
\newblock
\urldef\tempurl%
\url{https://www.usenix.org/conference/woot16/workshop-program/presentation/ispoglou}
\showURL{%
\tempurl}


\bibitem[\protect\citeauthoryear{Karbab, Debbabi, Derhab, and Mouheb}{Karbab
  et~al\mbox{.}}{2017}]%
        {android7}
\bibfield{author}{\bibinfo{person}{ElMouatez~Billah Karbab},
  \bibinfo{person}{Mourad Debbabi}, \bibinfo{person}{Abdelouahid Derhab}, {and}
  \bibinfo{person}{Djedjiga Mouheb}.} \bibinfo{year}{2017}\natexlab{}.
\newblock \showarticletitle{Android malware detection using deep learning on
  api method sequences}.
\newblock \bibinfo{journal}{\emph{arXiv preprint arXiv:1712.08996}}
  (\bibinfo{year}{2017}).
\newblock


\bibitem[\protect\citeauthoryear{Khormali, Abusnaina, Chen, Nyang, and
  Mohaisen}{Khormali et~al\mbox{.}}{2019}]%
        {pe1}
\bibfield{author}{\bibinfo{person}{Aminollah Khormali}, \bibinfo{person}{Ahmed
  Abusnaina}, \bibinfo{person}{Songqing Chen}, \bibinfo{person}{DaeHun Nyang},
  {and} \bibinfo{person}{Aziz Mohaisen}.} \bibinfo{year}{2019}\natexlab{}.
\newblock \showarticletitle{COPYCAT: Practical Adversarial Attacks on
  Visualization-Based Malware Detection}.
\newblock \bibinfo{journal}{\emph{arXiv preprint arXiv:1909.09735}}
  (\bibinfo{year}{2019}).
\newblock


\bibitem[\protect\citeauthoryear{Kolosnjaji, Demontis, Biggio, Maiorca,
  Giacinto, Eckert, and Roli}{Kolosnjaji et~al\mbox{.}}{2018}]%
        {pe12}
\bibfield{author}{\bibinfo{person}{Bojan Kolosnjaji}, \bibinfo{person}{Ambra
  Demontis}, \bibinfo{person}{Battista Biggio}, \bibinfo{person}{Davide
  Maiorca}, \bibinfo{person}{Giorgio Giacinto}, \bibinfo{person}{Claudia
  Eckert}, {and} \bibinfo{person}{Fabio Roli}.}
  \bibinfo{year}{2018}\natexlab{}.
\newblock \showarticletitle{Adversarial malware binaries: Evading deep learning
  for malware detection in executables}. In \bibinfo{booktitle}{\emph{2018 26th
  European Signal Processing Conference (EUSIPCO)}}. IEEE,
  \bibinfo{pages}{533--537}.
\newblock


\bibitem[\protect\citeauthoryear{Kouzemtchenko}{Kouzemtchenko}{2018}]%
        {android5}
\bibfield{author}{\bibinfo{person}{Alex Kouzemtchenko}.}
  \bibinfo{year}{2018}\natexlab{}.
\newblock \showarticletitle{Defending Malware Classification Networks Against
  Adversarial Perturbations with Non-Negative Weight Restrictions}.
\newblock \bibinfo{journal}{\emph{arXiv preprint arXiv:1806.09035}}
  (\bibinfo{year}{2018}).
\newblock


\bibitem[\protect\citeauthoryear{Kreuk, Barak, Aviv-Reuven, Baruch, Pinkas, and
  Keshet}{Kreuk et~al\mbox{.}}{2018a}]%
        {kreuk2018adversarial}
\bibfield{author}{\bibinfo{person}{Felix Kreuk}, \bibinfo{person}{Assi Barak},
  \bibinfo{person}{Shir Aviv-Reuven}, \bibinfo{person}{Moran Baruch},
  \bibinfo{person}{Benny Pinkas}, {and} \bibinfo{person}{Joseph Keshet}.}
  \bibinfo{year}{2018}\natexlab{a}.
\newblock \showarticletitle{Adversarial examples on discrete sequences for
  beating whole-binary malware detection}.
\newblock \bibinfo{journal}{\emph{arXiv preprint arXiv:1802.04528}}
  (\bibinfo{year}{2018}).
\newblock


\bibitem[\protect\citeauthoryear{Kreuk, Barak, Aviv-Reuven, Baruch, Pinkas, and
  Keshet}{Kreuk et~al\mbox{.}}{2018b}]%
        {pe13}
\bibfield{author}{\bibinfo{person}{Felix Kreuk}, \bibinfo{person}{Assi Barak},
  \bibinfo{person}{Shir Aviv-Reuven}, \bibinfo{person}{Moran Baruch},
  \bibinfo{person}{Benny Pinkas}, {and} \bibinfo{person}{Joseph Keshet}.}
  \bibinfo{year}{2018}\natexlab{b}.
\newblock \showarticletitle{Deceiving end-to-end deep learning malware
  detectors using adversarial examples}.
\newblock \bibinfo{journal}{\emph{arXiv preprint arXiv:1802.04528}}
  (\bibinfo{year}{2018}).
\newblock


\bibitem[\protect\citeauthoryear{Li, Baral, Li, Wang, Li, and Xu}{Li
  et~al\mbox{.}}{2018a}]%
        {android4}
\bibfield{author}{\bibinfo{person}{Deqiang Li}, \bibinfo{person}{Ramesh Baral},
  \bibinfo{person}{Tao Li}, \bibinfo{person}{Han Wang}, \bibinfo{person}{Qianmu
  Li}, {and} \bibinfo{person}{Shouhuai Xu}.} \bibinfo{year}{2018}\natexlab{a}.
\newblock \showarticletitle{Hashtran-dnn: A framework for enhancing robustness
  of deep neural networks against adversarial malware samples}.
\newblock \bibinfo{journal}{\emph{arXiv preprint arXiv:1809.06498}}
  (\bibinfo{year}{2018}).
\newblock


\bibitem[\protect\citeauthoryear{Li, Li, Ye, and Xu}{Li et~al\mbox{.}}{2018b}]%
        {pe6}
\bibfield{author}{\bibinfo{person}{Deqiang Li}, \bibinfo{person}{Qianmu Li},
  \bibinfo{person}{Yanfang Ye}, {and} \bibinfo{person}{Shouhuai Xu}.}
  \bibinfo{year}{2018}\natexlab{b}.
\newblock \showarticletitle{Enhancing Robustness of Deep Neural Networks
  Against Adversarial Malware Samples: Principles, Framework, and AICS'2019
  Challenge}.
\newblock \bibinfo{journal}{\emph{arXiv preprint arXiv:1812.08108}}
  (\bibinfo{year}{2018}).
\newblock


\bibitem[\protect\citeauthoryear{Lief}{Lief}{[n.d.]}]%
        {LIEF}
Lief \bibinfo{year}{[n.d.]}\natexlab{}.
\newblock \bibinfo{title}{{LIEF}}.
\newblock \bibinfo{howpublished}{\url{https://github.com/lief-project/LIEF}}.
\newblock


\bibitem[\protect\citeauthoryear{Liu, Du, Zhang, Zhu, Wang, and Guizani}{Liu
  et~al\mbox{.}}{2019a}]%
        {android3}
\bibfield{author}{\bibinfo{person}{Xiaolei Liu}, \bibinfo{person}{Xiaojiang
  Du}, \bibinfo{person}{Xiaosong Zhang}, \bibinfo{person}{Qingxin Zhu},
  \bibinfo{person}{Hao Wang}, {and} \bibinfo{person}{Mohsen Guizani}.}
  \bibinfo{year}{2019}\natexlab{a}.
\newblock \showarticletitle{Adversarial Samples on Android Malware Detection
  Systems for IoT Systems}.
\newblock \bibinfo{journal}{\emph{Sensors}} \bibinfo{volume}{19},
  \bibinfo{number}{4} (\bibinfo{year}{2019}), \bibinfo{pages}{974}.
\newblock


\bibitem[\protect\citeauthoryear{Liu, Zhang, Lin, and Li}{Liu
  et~al\mbox{.}}{2019b}]%
        {pe9}
\bibfield{author}{\bibinfo{person}{Xinbo Liu}, \bibinfo{person}{Jiliang Zhang},
  \bibinfo{person}{Yaping Lin}, {and} \bibinfo{person}{He Li}.}
  \bibinfo{year}{2019}\natexlab{b}.
\newblock \showarticletitle{Atmpa: Attacking machine learning-based malware
  visualization detection methods via adversarial examples}. In
  \bibinfo{booktitle}{\emph{2019 IEEE/ACM 27th International Symposium on
  Quality of Service (IWQoS)}}. IEEE, \bibinfo{pages}{1--10}.
\newblock


\bibitem[\protect\citeauthoryear{MAB}{MAB}{[n.d.]}]%
        {MAB}
MAB \bibinfo{year}{[n.d.]}\natexlab{}.
\newblock \bibinfo{title}{{Multi-armed bandit}}.
\newblock
  \bibinfo{howpublished}{\url{https://en.wikipedia.org/wiki/Multi-armed_bandit}}.
\newblock


\bibitem[\protect\citeauthoryear{Maiorca, Ariu, Corona, Aresu, and
  Giacinto}{Maiorca et~al\mbox{.}}{2015}]%
        {maiorca2015stealth}
\bibfield{author}{\bibinfo{person}{Davide Maiorca}, \bibinfo{person}{Davide
  Ariu}, \bibinfo{person}{Igino Corona}, \bibinfo{person}{Marco Aresu}, {and}
  \bibinfo{person}{Giorgio Giacinto}.} \bibinfo{year}{2015}\natexlab{}.
\newblock \showarticletitle{Stealth attacks: An extended insight into the
  obfuscation effects on android malware}.
\newblock \bibinfo{journal}{\emph{Computers \& Security}}  \bibinfo{volume}{51}
  (\bibinfo{year}{2015}), \bibinfo{pages}{16--31}.
\newblock


\bibitem[\protect\citeauthoryear{Maiorca, Biggio, Chiappe, and
  Giacinto}{Maiorca et~al\mbox{.}}{2017}]%
        {flash}
\bibfield{author}{\bibinfo{person}{Davide Maiorca}, \bibinfo{person}{Battista
  Biggio}, \bibinfo{person}{Maria~Elena Chiappe}, {and}
  \bibinfo{person}{Giorgio Giacinto}.} \bibinfo{year}{2017}\natexlab{}.
\newblock \showarticletitle{Adversarial Detection of Flash Malware: Limitations
  and Open Issues}.
\newblock \bibinfo{journal}{\emph{CoRR}}  \bibinfo{volume}{abs/1710.10225}
  (\bibinfo{year}{2017}).
\newblock


\bibitem[\protect\citeauthoryear{Maiorca, Biggio, and Giacinto}{Maiorca
  et~al\mbox{.}}{2018}]%
        {maiorca2018towards}
\bibfield{author}{\bibinfo{person}{Davide Maiorca}, \bibinfo{person}{Battista
  Biggio}, {and} \bibinfo{person}{Giorgio Giacinto}.}
  \bibinfo{year}{2018}\natexlab{}.
\newblock \showarticletitle{Towards Robust Detection of Adversarial Infection
  Vectors: Lessons Learned in PDF Malware}.
\newblock \bibinfo{journal}{\emph{arXiv preprint arXiv:1811.00830}}
  (\bibinfo{year}{2018}).
\newblock


\bibitem[\protect\citeauthoryear{Melis, Maiorca, Biggio, Giacinto, and
  Roli}{Melis et~al\mbox{.}}{2018}]%
        {android8}
\bibfield{author}{\bibinfo{person}{Marco Melis}, \bibinfo{person}{Davide
  Maiorca}, \bibinfo{person}{Battista Biggio}, \bibinfo{person}{Giorgio
  Giacinto}, {and} \bibinfo{person}{Fabio Roli}.}
  \bibinfo{year}{2018}\natexlab{}.
\newblock \showarticletitle{Explaining black-box android malware detection}. In
  \bibinfo{booktitle}{\emph{2018 26th European Signal Processing Conference
  (EUSIPCO)}}. IEEE, \bibinfo{pages}{524--528}.
\newblock


\bibitem[\protect\citeauthoryear{MLSEC2019}{MLSEC2019}{[n.d.]}]%
        {MLSEC2019}
MLSEC2019 \bibinfo{year}{[n.d.]}\natexlab{}.
\newblock \bibinfo{title}{{Machine Learning Static Evasion Competition 2019}}.
\newblock
  \bibinfo{howpublished}{\url{https://github.com/endgameinc/malware\_evasion\_competition}}.
\newblock


\bibitem[\protect\citeauthoryear{Papernot, McDaniel, Goodfellow, Jha, Celik,
  and Swami}{Papernot et~al\mbox{.}}{2017}]%
        {papernot2017practical}
\bibfield{author}{\bibinfo{person}{Nicolas Papernot}, \bibinfo{person}{Patrick
  McDaniel}, \bibinfo{person}{Ian Goodfellow}, \bibinfo{person}{Somesh Jha},
  \bibinfo{person}{Z~Berkay Celik}, {and} \bibinfo{person}{Ananthram Swami}.}
  \bibinfo{year}{2017}\natexlab{}.
\newblock \showarticletitle{Practical black-box attacks against machine
  learning}. In \bibinfo{booktitle}{\emph{Proceedings of the 2017 ACM on Asia
  conference on computer and communications security}}. ACM,
  \bibinfo{pages}{506--519}.
\newblock


\bibitem[\protect\citeauthoryear{Pappas, Polychronakis, and Keromytis}{Pappas
  et~al\mbox{.}}{2012}]%
        {ORP}
\bibfield{author}{\bibinfo{person}{Vasilis Pappas}, \bibinfo{person}{Michalis
  Polychronakis}, {and} \bibinfo{person}{Angelos~D Keromytis}.}
  \bibinfo{year}{2012}\natexlab{}.
\newblock \showarticletitle{Smashing the gadgets: Hindering return-oriented
  programming using in-place code randomization}. In
  \bibinfo{booktitle}{\emph{2012 IEEE Symposium on Security and Privacy}}.
  IEEE, \bibinfo{pages}{601--615}.
\newblock


\bibitem[\protect\citeauthoryear{Park, Khan, and Yener}{Park
  et~al\mbox{.}}{2019}]%
        {pe4}
\bibfield{author}{\bibinfo{person}{Daniel Park}, \bibinfo{person}{Haidar Khan},
  {and} \bibinfo{person}{B{\"{u}}lent Yener}.} \bibinfo{year}{2019}\natexlab{}.
\newblock \showarticletitle{Short Paper: Creating Adversarial Malware Examples
  using Code Insertion}.
\newblock \bibinfo{journal}{\emph{CoRR}}  \bibinfo{volume}{abs/1904.04802}
  (\bibinfo{year}{2019}).
\newblock


\bibitem[\protect\citeauthoryear{Pavithran, Patnaik, and Rebeiro}{Pavithran
  et~al\mbox{.}}{2019}]%
        {dtime}
\bibfield{author}{\bibinfo{person}{Jithin Pavithran}, \bibinfo{person}{Milan
  Patnaik}, {and} \bibinfo{person}{Chester Rebeiro}.}
  \bibinfo{year}{2019}\natexlab{}.
\newblock \showarticletitle{D-TIME: Distributed Threadless Independent Malware
  Execution for Runtime Obfuscation}. In \bibinfo{booktitle}{\emph{13th
  {USENIX} Workshop on Offensive Technologies ({WOOT} 19)}}.
  \bibinfo{publisher}{{USENIX} Association}, \bibinfo{address}{Santa Clara,
  CA}.
\newblock
\urldef\tempurl%
\url{https://www.usenix.org/conference/woot19/presentation/pavithran}
\showURL{%
\tempurl}


\bibitem[\protect\citeauthoryear{pcmag}{pcmag}{2020}]%
        {pcmag}
pcmag \bibinfo{year}{2020}\natexlab{}.
\newblock \bibinfo{title}{The best antivirus protection}.
\newblock
  \bibinfo{howpublished}{\url{https://www.pcmag.com/picks/the-best-antivirus-protection}}.
\newblock


\bibitem[\protect\citeauthoryear{Pierazzi, Pendlebury, Cortellazzi, and
  Cavallaro}{Pierazzi et~al\mbox{.}}{2020}]%
        {pierazzi2019intriguing}
\bibfield{author}{\bibinfo{person}{Fabio Pierazzi}, \bibinfo{person}{Feargus
  Pendlebury}, \bibinfo{person}{Jacopo Cortellazzi}, {and}
  \bibinfo{person}{Lorenzo Cavallaro}.} \bibinfo{year}{2020}\natexlab{}.
\newblock \showarticletitle{Intriguing Properties of Adversarial ML Attacks in
  the Problem Space}.
\newblock \bibinfo{journal}{\emph{2020 IEEE Security and Privacy}}
  (\bibinfo{year}{2020}).
\newblock


\bibitem[\protect\citeauthoryear{Podschwadt and Takabi}{Podschwadt and
  Takabi}{2019}]%
        {android1}
\bibfield{author}{\bibinfo{person}{Robert Podschwadt} {and}
  \bibinfo{person}{Hassan Takabi}.} \bibinfo{year}{2019}\natexlab{}.
\newblock \showarticletitle{Effectiveness of Adversarial Examples and Defenses
  for Malware Classification}.
\newblock \bibinfo{journal}{\emph{arXiv preprint arXiv:1909.04778}}
  (\bibinfo{year}{2019}).
\newblock


\bibitem[\protect\citeauthoryear{Quiring, Maier, and Rieck}{Quiring
  et~al\mbox{.}}{2019}]%
        {quiring2019misleading}
\bibfield{author}{\bibinfo{person}{Erwin Quiring}, \bibinfo{person}{Alwin
  Maier}, {and} \bibinfo{person}{Konrad Rieck}.}
  \bibinfo{year}{2019}\natexlab{}.
\newblock \showarticletitle{Misleading authorship attribution of source code
  using adversarial learning}. In \bibinfo{booktitle}{\emph{28th $\{$USENIX$\}$
  Security Symposium ($\{$USENIX$\}$ Security 19)}}. \bibinfo{pages}{479--496}.
\newblock


\bibitem[\protect\citeauthoryear{Raff, Barker, Sylvester, Brandon, Catanzaro,
  and Nicholas}{Raff et~al\mbox{.}}{2018}]%
        {malconv}
\bibfield{author}{\bibinfo{person}{Edward Raff}, \bibinfo{person}{Jon Barker},
  \bibinfo{person}{Jared Sylvester}, \bibinfo{person}{Robert Brandon},
  \bibinfo{person}{Bryan Catanzaro}, {and} \bibinfo{person}{Charles~K
  Nicholas}.} \bibinfo{year}{2018}\natexlab{}.
\newblock \showarticletitle{Malware detection by eating a whole exe}. In
  \bibinfo{booktitle}{\emph{Workshops at the Thirty-Second AAAI Conference on
  Artificial Intelligence}}.
\newblock


\bibitem[\protect\citeauthoryear{Rieck, Trinius, Willems, and Holz}{Rieck
  et~al\mbox{.}}{2011}]%
        {rieck2011automatic}
\bibfield{author}{\bibinfo{person}{Konrad Rieck}, \bibinfo{person}{Philipp
  Trinius}, \bibinfo{person}{Carsten Willems}, {and} \bibinfo{person}{Thorsten
  Holz}.} \bibinfo{year}{2011}\natexlab{}.
\newblock \showarticletitle{Automatic analysis of malware behavior using
  machine learning}.
\newblock \bibinfo{journal}{\emph{Journal of Computer Security}}
  \bibinfo{volume}{19}, \bibinfo{number}{4} (\bibinfo{year}{2011}),
  \bibinfo{pages}{639--668}.
\newblock


\bibitem[\protect\citeauthoryear{Rosenberg, Shabtai, Elovici, and
  Rokach}{Rosenberg et~al\mbox{.}}{2018a}]%
        {pe10}
\bibfield{author}{\bibinfo{person}{Ishai Rosenberg}, \bibinfo{person}{Asaf
  Shabtai}, \bibinfo{person}{Yuval Elovici}, {and} \bibinfo{person}{Lior
  Rokach}.} \bibinfo{year}{2018}\natexlab{a}.
\newblock \showarticletitle{Query-Efficient GAN Based Black-Box Attack Against
  Sequence Based Machine and Deep Learning Classifiers}.
\newblock \bibinfo{journal}{\emph{arXiv preprint arXiv:1804.08778}}
  (\bibinfo{year}{2018}).
\newblock


\bibitem[\protect\citeauthoryear{Rosenberg, Shabtai, Rokach, and
  Elovici}{Rosenberg et~al\mbox{.}}{2018b}]%
        {pe17}
\bibfield{author}{\bibinfo{person}{Ishai Rosenberg}, \bibinfo{person}{Asaf
  Shabtai}, \bibinfo{person}{Lior Rokach}, {and} \bibinfo{person}{Yuval
  Elovici}.} \bibinfo{year}{2018}\natexlab{b}.
\newblock \showarticletitle{Generic black-box end-to-end attack against state
  of the art API call based malware classifiers}. In
  \bibinfo{booktitle}{\emph{International Symposium on Research in Attacks,
  Intrusions, and Defenses}}. Springer, \bibinfo{pages}{490--510}.
\newblock


\bibitem[\protect\citeauthoryear{Rosenberg, Shabtai, Rokach, and
  Elovici}{Rosenberg et~al\mbox{.}}{2018c}]%
        {ISRAID2018genericblackbox}
\bibfield{author}{\bibinfo{person}{Ishai Rosenberg}, \bibinfo{person}{Asaf
  Shabtai}, \bibinfo{person}{Lior Rokach}, {and} \bibinfo{person}{Yuval
  Elovici}.} \bibinfo{year}{2018}\natexlab{c}.
\newblock \showarticletitle{Generic black-box end-to-end attack against state
  of the art API call based malware classifiers}. In
  \bibinfo{booktitle}{\emph{International Symposium on Research in Attacks,
  Intrusions, and Defenses}}. Springer, \bibinfo{pages}{490--510}.
\newblock


\bibitem[\protect\citeauthoryear{Saxe and Berlin}{Saxe and Berlin}{2015}]%
        {saxe2015deep}
\bibfield{author}{\bibinfo{person}{Joshua Saxe} {and}
  \bibinfo{person}{Konstantin Berlin}.} \bibinfo{year}{2015}\natexlab{}.
\newblock \showarticletitle{Deep neural network based malware detection using
  two dimensional binary program features}. In \bibinfo{booktitle}{\emph{2015
  10th International Conference on Malicious and Unwanted Software (MALWARE)}}.
  IEEE, \bibinfo{pages}{11--20}.
\newblock


\bibitem[\protect\citeauthoryear{Schultz, Eskin, Zadok, and Stolfo}{Schultz
  et~al\mbox{.}}{2000}]%
        {schultz2000data}
\bibfield{author}{\bibinfo{person}{Matthew~G Schultz}, \bibinfo{person}{Eleazar
  Eskin}, \bibinfo{person}{F Zadok}, {and} \bibinfo{person}{Salvatore~J
  Stolfo}.} \bibinfo{year}{2000}\natexlab{}.
\newblock \showarticletitle{Data mining methods for detection of new malicious
  executables}. In \bibinfo{booktitle}{\emph{Proceedings 2001 IEEE Symposium on
  Security and Privacy. S\&P 2001}}. IEEE, \bibinfo{pages}{38--49}.
\newblock


\bibitem[\protect\citeauthoryear{Stokes, Wang, Marinescu, Marino, and
  Bussone}{Stokes et~al\mbox{.}}{2017}]%
        {stokes2017attack}
\bibfield{author}{\bibinfo{person}{Jack~W Stokes}, \bibinfo{person}{De Wang},
  \bibinfo{person}{Mady Marinescu}, \bibinfo{person}{Marc Marino}, {and}
  \bibinfo{person}{Brian Bussone}.} \bibinfo{year}{2017}\natexlab{}.
\newblock \showarticletitle{Attack and defense of dynamic analysis-based,
  adversarial neural malware classification models}.
\newblock \bibinfo{journal}{\emph{arXiv preprint arXiv:1712.05919}}
  (\bibinfo{year}{2017}).
\newblock


\bibitem[\protect\citeauthoryear{Stokes, Wang, Marinescu, Marino, and
  Bussone}{Stokes et~al\mbox{.}}{2018}]%
        {pe16}
\bibfield{author}{\bibinfo{person}{Jack~W Stokes}, \bibinfo{person}{De Wang},
  \bibinfo{person}{Mady Marinescu}, \bibinfo{person}{Marc Marino}, {and}
  \bibinfo{person}{Brian Bussone}.} \bibinfo{year}{2018}\natexlab{}.
\newblock \showarticletitle{Attack and Defense of Dynamic Analysis-Based,
  Adversarial Neural Malware Detection Models}. In
  \bibinfo{booktitle}{\emph{MILCOM 2018-2018 IEEE Military Communications
  Conference (MILCOM)}}. IEEE, \bibinfo{pages}{1--8}.
\newblock


\bibitem[\protect\citeauthoryear{Suciu, Coull, and Johns}{Suciu
  et~al\mbox{.}}{2019}]%
        {pe7}
\bibfield{author}{\bibinfo{person}{Octavian Suciu}, \bibinfo{person}{Scott~E
  Coull}, {and} \bibinfo{person}{Jeffrey Johns}.}
  \bibinfo{year}{2019}\natexlab{}.
\newblock \showarticletitle{Exploring adversarial examples in malware
  detection}. In \bibinfo{booktitle}{\emph{2019 IEEE Security and Privacy
  Workshops (SPW)}}. IEEE, \bibinfo{pages}{8--14}.
\newblock


\bibitem[\protect\citeauthoryear{Taheri, Javidan, Shojafar, Pooranian, Miri,
  and Conti}{Taheri et~al\mbox{.}}{2019}]%
        {android2}
\bibfield{author}{\bibinfo{person}{Rahim Taheri}, \bibinfo{person}{Reza
  Javidan}, \bibinfo{person}{Mohammad Shojafar}, \bibinfo{person}{Zahra
  Pooranian}, \bibinfo{person}{Ali Miri}, {and} \bibinfo{person}{Mauro Conti}.}
  \bibinfo{year}{2019}\natexlab{}.
\newblock \showarticletitle{On Defending Against Label Flipping Attacks on
  Malware Detection Systems}.
\newblock \bibinfo{journal}{\emph{arXiv preprint arXiv:1908.04473}}
  (\bibinfo{year}{2019}).
\newblock


\bibitem[\protect\citeauthoryear{Team}{Team}{2019}]%
        {microsoft_link}
\bibfield{author}{\bibinfo{person}{Microsoft Defender ATP~Research Team}.}
  \bibinfo{year}{2019}\natexlab{}.
\newblock \bibinfo{title}{New machine learning model sifts through the good to
  unearth the bad in evasive malware}.
\newblock
  \bibinfo{howpublished}{\url{https://www.microsoft.com/security/blog/2019/07/25/new-machine-learning-model-sifts-through-the-good-to-unearth-the-bad-in-evasive-malware/}}.
\newblock


\bibitem[\protect\citeauthoryear{TS}{TS}{[n.d.]}]%
        {TS}
TS \bibinfo{year}{[n.d.]}\natexlab{}.
\newblock \bibinfo{title}{{Thompson Sampling}}.
\newblock
  \bibinfo{howpublished}{\url{https://en.wikipedia.org/wiki/Thompson_sampling}}.
\newblock


\bibitem[\protect\citeauthoryear{Xu, Qi, and Evans}{Xu et~al\mbox{.}}{2016}]%
        {xu2016automatically}
\bibfield{author}{\bibinfo{person}{Weilin Xu}, \bibinfo{person}{Yanjun Qi},
  {and} \bibinfo{person}{David Evans}.} \bibinfo{year}{2016}\natexlab{}.
\newblock \showarticletitle{Automatically evading classifiers}. In
  \bibinfo{booktitle}{\emph{Proceedings of the 2016 network and distributed
  systems symposium}}. \bibinfo{pages}{21--24}.
\newblock


\bibitem[\protect\citeauthoryear{Yang, Kong, Xie, and Gunter}{Yang
  et~al\mbox{.}}{2017}]%
        {yang2017malware}
\bibfield{author}{\bibinfo{person}{Wei Yang}, \bibinfo{person}{Deguang Kong},
  \bibinfo{person}{Tao Xie}, {and} \bibinfo{person}{Carl~A Gunter}.}
  \bibinfo{year}{2017}\natexlab{}.
\newblock \showarticletitle{Malware detection in adversarial settings:
  Exploiting feature evolutions and confusions in android apps}. In
  \bibinfo{booktitle}{\emph{Proceedings of the 33rd Annual Computer Security
  Applications Conference}}. \bibinfo{pages}{288--302}.
\newblock


\end{thebibliography}

\appendix
\section{Broken Malware Examples}
\label{appendix:broken_malware_examples}

\begin{figure}[ht]
    \centering
    \includegraphics[width=0.8\linewidth]{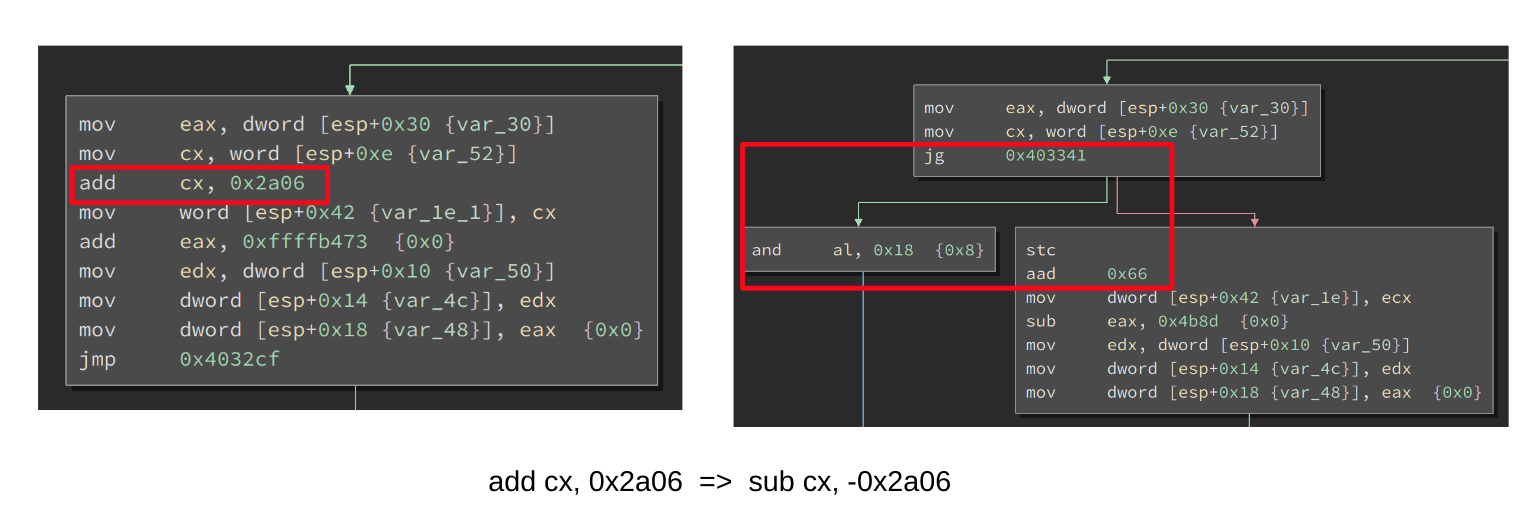}
    \caption{Implementation Errors in CR}
    \label{fig:break_CR1}
\end{figure}

Case 1: Implementation errors in instruction replacement. As shown in Figure~\ref{fig:break_CR1}, the original implementation of code randomization only supports 8-bit and 32-bit, not 16-bit instruction. It tries to replace a 16-bit add/sub instruction in a wrong way (assuming 32-bit format). It treats the last four bytes as the second operand, but actually, only the last two bytes are the second operand. This would break the CFG of the program.

\begin{figure}[ht]
    \centering
    \includegraphics[width=0.8\linewidth]{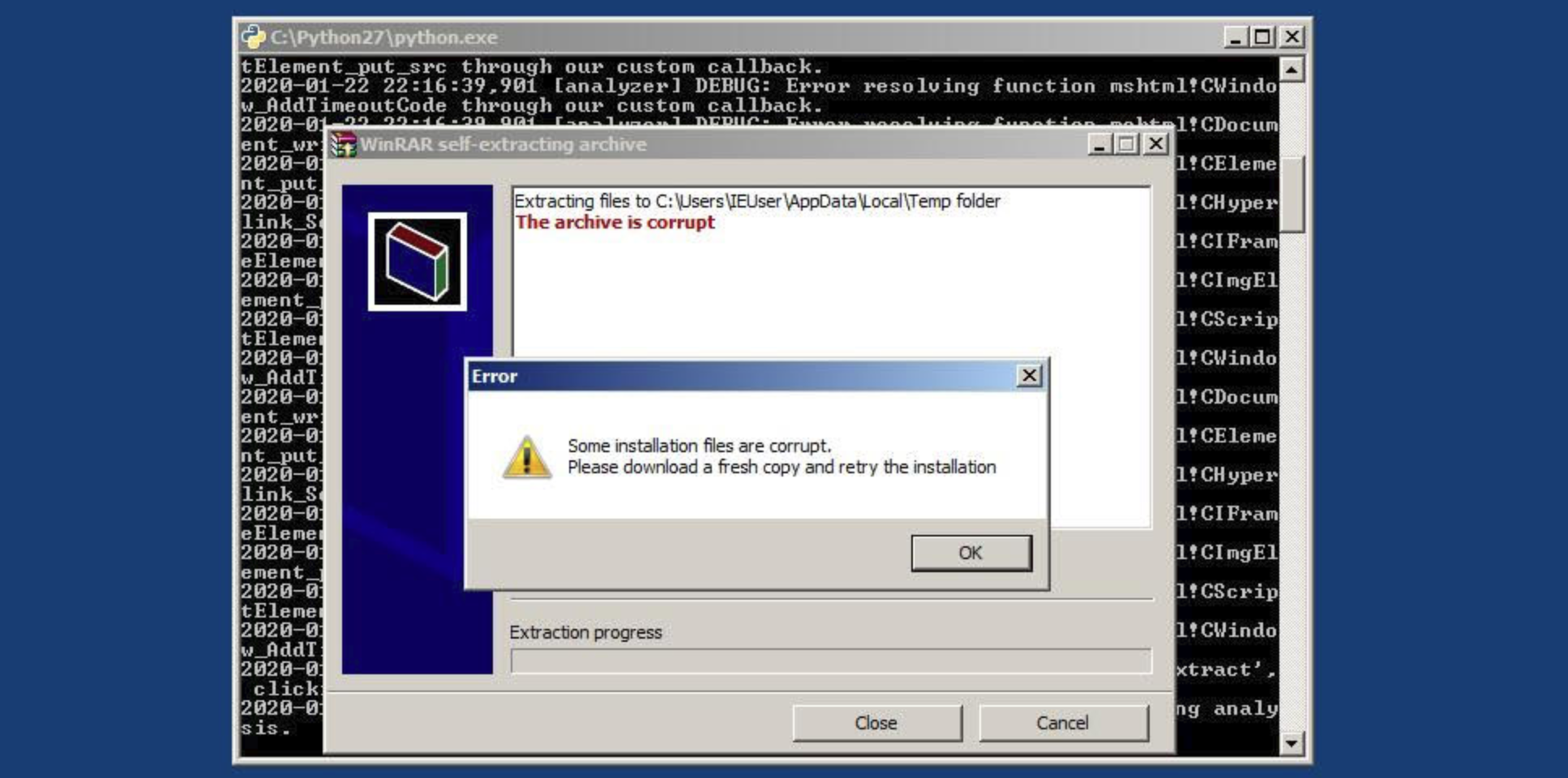}
    \caption{Errors after Overlay Data Append}
    \label{fig:break2}
\end{figure}

\medskip\noindent Case 2: Overwriting overlay. As shown in Figure~\ref{fig:break2}, when adding a new section at the end of the last section, if the sample has overlay data, the added new data may affect the overlay data extraction of the malware.

\textbf{}



\end{document}